\documentclass[12pt]{iopart}

\usepackage{amssymb,bm,epsfig,color,subfig}
\usepackage{iopams}

\usepackage{amsfonts}
\usepackage[english]{babel}
\usepackage{latexsym}
\usepackage{graphics}
\usepackage{epsfig}
\usepackage{dcolumn}
\usepackage{epstopdf}
\usepackage{hyperref}
\usepackage{multirow}
\usepackage{float}
\usepackage{times}

\usepackage{verbatim}

\begin{document}

\title[Frequentist and Bayesian Quantum Phase Estimation]{Frequentist and Bayesian Quantum Phase Estimation}

\author{Y. Li $^1$\footnote{These authors contributed equally to this work}, 
L. Pezz\`e $^2$$\ddagger$, M. Gessner $^2$$\ddagger$, W. Li $^1$\footnote{Corresponding author, wdli@sxu.edu.cn}, and A. Smerzi $^2$}
\address{1 Institute of Theoretical Physics and Department of Physics, Shanxi University, 030006, Taiyuan, China}
\address{2 QSTAR, INO-CNR and LENS, Largo Enrico Fermi 2, 50125 Firenze, Italy}

\vspace{10pt}
\begin{indented}
\item[]\today
\end{indented}

\begin{abstract}
Frequentist and Bayesian phase estimation strategies lead to conceptually different results on the state of knowledge about the true value of the phase shift. 
We compare the two frameworks and their sensitivity bounds to the estimation of an interferometric phase shift limited by quantum noise, 
considering both the cases of a fixed and a fluctuating parameter.
We point out that frequentist precision bounds, such as the Cram\`er-Rao bound, for instance, do not apply to Bayesian strategies and vice-versa. 
Similarly, bounds for fluctuating parameters make no statement about the estimation of a fixed parameter.
\end{abstract}




\section{Introduction}

The estimation of a phase shift using interferometric techniques is at the core of metrology and sensing \cite{MZ, Ramsey}.
Applications range from the definition of the standard of time \cite{Clocks2015}
to the detection of gravitational waves \cite{GravWaves2013}.
The general problem can be concisely stated as the search for optimal strategies to
minimize the phase estimation uncertainty.
The noise that limits the achievable phase sensitivity can have a ``classical'' or a ``quantum'' nature.
Classical noise originates from the coupling of the interferometer with some external source of disturbance, like seismic vibrations, parasitic magnetic fields or
from incoherent interactions within the interferometer. Such noise can, in principle, be arbitrarily reduced, e.g., by shielding the interferometer from external
noise or by tuning interaction parameters to ensure a fully coherent time evolution.
The second source of uncertainty has an irreducible quantum origin \cite{HelstromBOOK1976}.
Quantum noise cannot be fully suppressed, even in the idealized case of the creation and manipulation of pure quantum states.
Using classically-correlated probe states it is possible to reach the so-called shot noise or standard quantum limit,
which is the limiting factor for the current generation of interferometers and sensors \cite{LudlowRMP2015, SchnabelNATCOMM2013, Cronin2009}.
Strategies involving probe states characterized by squeezed quadratures \cite{CavesPRD1981}
or entanglement between particles \cite{GovannettiPRL2006, PezzePRL2009, Varenna, Toth} are able to overcome the shot noise, the ultimate quantum bound being the so-called
Heisenberg limit. Quantum noise reduction in phase estimation has been demonstrated in several proof-of-principle experiments
with atoms and photons \cite{Giovannetti2011, PezzeRMP}.

There is a vast literature dealing with the parameter estimation problem which has been mostly developed following
two different approaches \cite{KayBook, Lehmann, VanTreesBook2007}: frequentist and Bayesian.
Both approaches have been investigated in the context of quantum phase estimation
\cite{Varenna, Giovannetti2011, LanePRA1993, Tsang2012, QWWB, HallNJP2012, GiovannettiPRL2012, PezzePRA2013, PezzePRA2015}
and implemented/tested experimentally \cite{Hradil, PezzePRL2007, KacprowiczNATPHOT2010, KrischekPRL2011, XiangNATPHOT201}.
They build on conceptually different meanings
attached to the word ``probability", and their respective results provide
conceptually different information on the estimated parameters and their uncertainties.

In the limit of a large number of repeated measurements, the sensitivity reached by the 
frequentist and Bayesian methods often asymptotically agree: this fact has very often induced to believe that
the two paradigms can be interchangeably used in the phase estimation theory without acknowledging their irreconcilable nature.
Overlooking these differences is not only conceptually inconsistent but can even create paradoxes,
as, for instance, the existence of ultimate bounds in sensitivity proven in one paradigm that can be violated in the other.

In this manuscript we directly compare the frequentist and the Bayesian parameter estimation theory.
We study different sensitivity bounds obtained in the two frameworks and highlight the conceptual differences
between the two. 
Besides the asymptotic regime of many repeated measurements, we also study bounds that are relevant for small samples.

Our results are illustrated with a simple test model \cite{BollingerPRA1996, PezzeEPL2007}. 
We consider $N$ qubits with basis states $|0\rangle$ and $|1\rangle$, initially prepared in
a (generalized) GHZ state $\vert \mathrm{GHZ} \rangle = (|0\rangle^{\otimes N}+|1\rangle^{\otimes N})/\sqrt{2}$,
with all particles being either in $|1\rangle$ or in $|0\rangle$. The phase-encoding is a rotation of each qubit in the Bloch sphere
$|0\rangle \to e^{-i \theta/2} |0\rangle$ and $|1\rangle \to e^{+i \theta/2} |1\rangle$,
which transforms the $\vert \mathrm{GHZ} \rangle$ state into
$\vert \mathrm{GHZ} (\theta) \rangle =  (e^{-i N \theta/2} |0\rangle^{\otimes N}+ e^{+i N \theta/2} |1\rangle^{\otimes N}) / \sqrt{2}$.
The phase is estimated by measuring the parity $(-1)^{N_0}$,
where $N_0=0,1$ is the number of particles in the state $|0\rangle$ \cite{BollingerPRA1996, GerryCP2010, Wineland2000, Monz2011}.
The parity measurement has two possible results $\mu = \pm1$ that are conditioned by the ``true value of the phase
shift" $\theta_0$ with probability $p(\pm1|\theta_0)=(1\pm\cos\left(  N \theta_0\right) )/2$.
The probability to observe the sequence of results
$\bm{\mu} = \{\mu_1,\mu_2,\dots, \mu_m\}$ in $m$ independent repetitions of the experiment (with same probe state and phase encoding transformation) is
\begin{equation}
p(\bm{\mu}|\theta_0)={\prod_{i=1}^m}p(\mu_{i}|\theta_0)=\left(  \frac{1+\cos\left(  N\theta_0\right)  }{2}\right)
^{m_+}\left(  \frac{1-\cos\left(  N\theta_0\right)  }{2}\right)  ^{m_-}, \label{mlf}
\end{equation}
where $m_{\pm}$ is the number of the observed results $\pm 1$, respectively. Notice that $p(\bm{\mu}|\theta_0)$ 
is the conditional probability for the measurement outcome $\bm{\mu}$,
given that the true value of the phase shift is $\theta_0$ (which we consider to be unknown in the estimation protocol).
Equation~(\ref{mlf}) provides the probability that will be used in the following sections for the case $N=2$.
Sections~\ref{sec:freq} and~\ref{sec:Bayes} deal with the case where $\theta_0$ has a fixed value and in Section~\ref{sec:rand} 
we discuss precision bounds for a fluctuating phase shift.

\section{Frequentist approach}\label{sec:freq}

In the frequentist paradigm, the phase (assumed having a fixed but unknown value $\theta_0$) is estimated via an
arbitrarily chosen function of the measurement results, $\theta_{\mathrm{est}}(\bm{\mu})$, called the estimator.
Typically, $\theta_{\mathrm{est}}(\bm{\mu})$ is chosen by maximizing the likelihood of the observed data (see below).
The estimator, being a function of random outcomes, is itself a random variable.
It is characterized by a statistical distribution that has an objective, measurable character.
The relative frequency with which the event $\theta_{\mathrm{est}}$ occurs 
converges to a probability asymptotically with the number of repeated experimental trials.

\subsection{Frequentist risk functions}
Statistical fluctuations of the data reflect the {\it statistical uncertainty} of the estimation.
This is quantified by the variance,
\begin{equation} \label{var}
\big( \Delta^2 \theta_{\rm est} \big)_{\bm{\mu} \vert \theta_0}=
\sum_{\bm{\mu}} \big( \theta_{\mathrm{est}}(\bm{\mu}) - \langle \theta_{\mathrm{est}} \rangle_{\bm{\mu} \vert \theta_0} \big)^2 p(\bm{\mu} \vert \theta_0),
\end{equation}
around the mean value $\langle \theta_{\mathrm{est}} \rangle_{\bm{\mu} \vert \theta_0}
=  \sum_{\bm{\mu}} \theta_{\mathrm{est}}(\bm{\mu}) p(\bm{\mu} \vert \theta_0)$,
 the sum extending over all possible measurement sequences (for fixed $\theta_0$ and $m$).
An important class is that of {\it locally unbiased} estimators, namely those satisfying $\langle \theta_{\mathrm{est}} \rangle_{\bm{\mu} \vert \theta_0} = \theta_0$ 
and $\frac{ d \langle \theta_{\mathrm{est}} \rangle_{\bm{\mu} \vert \theta} }{ d \theta } \big\vert_{\theta=\theta_0}=1$, see for instance \cite{Hayashi}.
An estimator is unbiased if and only if it is locally unbiased at every $\theta_0$.

The quality of the estimator can also be quantified by mean square error (MSE)~\cite{Lehmann}
\begin{equation} \label{MSE}
\mathrm{MSE}(\theta_{\mathrm{est}})_{\bm{\mu}|\theta_0}=
\sum_{\bm{\mu}} \big( \theta_{\mathrm{est}}(\bm{\mu}) - \theta_0 \big)^2 p(\bm{\mu} \vert \theta_0),
\end{equation}
giving the deviation of $\theta_{\mathrm{est}}$ from the true value of the phase shift $\theta_0$.
It is related to Eq.~(\ref{var}) by the relation
\begin{equation} \label{varMSEbias}
\mathrm{MSE}(\theta_{\mathrm{est}})_{\bm{\mu}|\theta_0}=\big( \Delta^2 \theta_{\rm est} \big)_{\bm{\mu} \vert \theta_0} + \left(\langle\theta_{\mathrm{est}}\rangle_{\boldsymbol{\mu}|\theta_0}-\theta_0\right)^2.
\end{equation}
Notice that the MSE cannot be accessed from the experimentally available data since the true value $\theta_0$ is unknown.
In this sense, only the fluctuations of $\theta_{\mathrm{est}}$ around its mean value, i.e., the variance $(\Delta^2 \theta_{\rm est})_{\bm{\mu} \vert \theta_0}$, have experimental relevance.
For unbiased estimators, Eqs.~(\ref{var}) and (\ref{varMSEbias}) coincide.
In general, since the bias term in Eq.~(\ref{varMSEbias}) is never negative, $\mathrm{MSE}(\theta_{\mathrm{est}})_{\bm{\mu}|\theta_0}\geq \big( \Delta^2 \theta_{\rm est} \big)_{\bm{\mu} \vert \theta_0}$ and
any lower bound on $( \Delta^2 \theta_{\rm est})_{\bm{\mu} \vert \theta_0}$ automatically provides a lower bound on $\mathrm{MSE}(\theta_{\mathrm{est}})_{\bm{\mu}|\theta_0}$ but not vice-versa.
In the following section, we therefore limit our attention to bounds on $( \Delta^2 \theta_{\rm est})_{\bm{\mu} \vert \theta_0}$.
The distinction between the two quantities becomes more important in the case of a fluctuating phase shift $\theta_0$,
where the bias can affect the corresponding bounds in different ways. We will see this explicitly in Sec.~\ref{sec:rand}.

\subsection{Frequentist bounds on phase sensitivity}

\subsubsection{Barankin bound.}

The Barankin bound (BB) provides the tightest lower bound to the variance~(\ref{var})~\cite{Barankin1949}.
It can be proven to be always (for any $m$) saturable, in principle,
by a specific local (i.e., dependent of $\theta_0$) estimator and measurement observable.
The BB can be written as \cite{Mcaulay1971}
\begin{equation}
\left( \Delta^2 \theta_{\mathrm{est}} \right)_{\bm{\mu}|\theta_0} \geq \Delta^2 \theta_{\rm BB} \equiv \sup_{\theta_i,a_i,n}
\frac{\left\{\sum_{i=1}^{n}a_i[ \langle \theta_{\mathrm{est}} \rangle_{\bm{\mu} \vert \theta_i} - \langle \theta_{\mathrm{est}} \rangle_{\bm{\mu} \vert \theta_0} ]\right\}^2}{
\sum_{\bm{\mu}} \left[\sum_{i=1}^{n}a_i L(\bm{\mu} \vert \theta_i,\theta_0)\right]^2 p(\bm{\mu}|\theta_0)},  \label{BB}
\end{equation}
where $L(\bm{\mu} \vert \theta_i,\theta)=p(\bm{\mu}|\theta_i)/p(\bm{\mu}|\theta)$ is generally indicated as likelihood ratio
and the supremum is taken over $n$ parameters $a_i \in \mathbb{R}$, which are arbitrary real numbers, and $\theta_i$, which are arbitrary phase values in the parameter domain.
For unbiased estimators, we can replace $\langle \theta_{\mathrm{est}} \rangle_{\bm{\mu} \vert \theta_i}=\theta_i$ for all $i$ and the BB becomes independent of the estimator:
\begin{equation}
\left( \Delta^2 \theta_{\mathrm{est}} \right)_{\bm{\mu}|\theta_0} \geq \Delta^2 \theta_{\rm BB}^{\mathrm{ub}} \equiv \sup_{\theta_i,a_i,n} \frac{\left\{\sum_{i=1}^{n}a_i[\theta_i - \theta_0 ]\right\}^2}{
\sum_{\bm{\mu}} \left[\sum_{i=1}^{n}a_i L(\bm{\mu} \vert \theta_i,\theta_0)\right]^2 p(\bm{\mu}|\theta_0)}.  \label{BBub}
\end{equation}
A derivation of the BB is presented in \ref{app:Barankin}.

The explicit calculation of $\Delta^2 \theta_{\rm BB}$ is impractical in most applications
due to the number of free variables that must be optimized.
However, the BB provides a strict hierarchy of bounds of increasing complexity that can be of great practical importance.
Restricting the number of variables in the optimization can provide local lower bounds that are much simpler to determine
at the expense of not being saturable in general, namely, for an arbitrary number of measurements.
Below, we demonstrate the following hierarchy of bounds:
\begin{equation} \label{ineq}
\left( \Delta^2 \theta_{\mathrm{est}} \right)_{\bm{\mu}|\theta_0}
\geq \Delta^2 \theta_{\rm BB} \geq \Delta^2 \theta_{\rm EChRB} \geq \Delta^2 \theta_{\rm ChRB} \geq
\Delta^2 \theta_{\rm CRLB},
\end{equation}
where $\Delta^2 \theta_{\rm CRLB}$ is the Cram\'{e}r-Rao lower bound (CRLB) \cite{Cramer1946, Rao1945}
and $\Delta^2 \theta_{\rm ChRB}$ is the Hammersley-Chapman-Robbins bound (ChRB) \cite{Hammersley1950,Chapman1951}.
We will also introduce a novel extended version of the ChRB, indicated as $\Delta^2 \theta_{\rm EChRB}$.

\subsubsection{Cram\'{e}r-Rao lower bound and maximum likelihood estimator.}

The CRLB is the most common frequentist bound in parameter estimation.
It is given by \cite{Cramer1946, Rao1945}:
\begin{equation} \label{CRLB}
\Delta^2 \theta_{\rm CRLB} = \frac{ \left( \frac{d \langle \theta_{\mathrm{est}} \rangle_{\bm{\mu} \vert \theta_0} }{d\theta_0} \right)^2}{m F(\theta_0)}.
\end{equation}
The inequality $\left( \Delta^2 \theta_{\mathrm{est}} \right)_{\bm{\mu}|\theta_0} \geq \Delta^2 \theta_{\rm CRLB}$
is obtained by differentiating  $\langle \theta_{\mathrm{est}} \rangle_{\bm{\mu} \vert \theta_0}$
with respect to $\theta_0$ and using a Cauchy-Schwarz inequality:
\begin{equation}
\fl \bigg( \frac{d \langle \theta_{\mathrm{est}} \rangle_{\bm{\mu} \vert \theta_0} }{d\theta_0} \bigg)^2
= \bigg( \sum_{\bm{\mu}} \big( \theta_{\mathrm{est}}(\bm{\mu}) -\langle \theta_{\mathrm{est}}
\rangle_{\bm{\mu} \vert \theta_0} \big) \frac{d p(\bm{\mu} \vert \theta_0)}{d \theta_0} \bigg)^2 \leq m F(\theta_0)
\big( \Delta^2 \theta_{\rm est} \big)_{\bm{\mu} \vert \theta_0},
\end{equation}
where we have used $\sum_{\bm{\mu}} \frac{d p(\bm{\mu} \vert \theta_0)}{d \theta_0} = 0$ and
$\sum_{\bm{\mu}} \frac{1}{p(\bm{\mu}|\theta_0)}
(  \frac{\partial p(\bm{\mu}|\theta)}{\partial\theta} \vert_{\theta_0} )  ^{2}
= m \sum_{\mu}
\frac{1}{p(\mu|\theta_0)}
(  \frac{\partial p(\mu|\theta)}{\partial\theta} \vert_{\theta_0} )  ^{2}$ valid for $m$ independent measurements, and
\begin{equation}
F\left(  \theta_0 \right) = \sum_{\mu}
\frac{1}{p(\mu|\theta_0)}
\left(  \frac{\partial p(\mu|\theta)}{\partial\theta} \Big\vert_{\theta_0} \right)  ^{2} \label{fisher}
\end{equation}
is the Fisher information.
The equality $\left( \Delta^2 \theta_{\mathrm{est}} \right)_{\bm{\mu}|\theta_0} = \Delta^2 \theta_{\rm CRLB}$ is achieved if and only if
\begin{equation}
\theta_{\rm est}(\bm{\mu}) - \langle \theta_{\rm est} \rangle_{\bm{\mu} \vert \theta_0} = \lambda_{\theta_0} \frac{d \log p(\bm{\mu}\vert \theta_0)}{d\theta_0},
\end{equation}
with $\lambda_{\theta_0}$ a parameter independent of $\bm{\mu}$ (while it may depend on $\theta_0$).
Noticing that
$ \frac{d \langle \theta_{\mathrm{est}} \rangle_{\bm{\mu} \vert \theta_0} }{d\theta_0}  =
\sum_{\bm{\mu}} \big( \theta_{\mathrm{est}}(\bm{\mu}) - f(\theta_0) \big) \frac{d p(\bm{\mu} \vert \theta_0)}{d \theta_0}$,
the CRLB can be straightforwardly generalized to any function $f(\theta_0)$ independent of $\bm{\mu}$.
In particular, choosing $f(\theta_0) = \theta_0$, we can directly prove that $\mathrm{MSE}(\theta_{\mathrm{est}})_{\bm{\mu}|\theta_0} \geq \Delta^2 \theta_{\rm CRLB}$,
which also depends on the bias.

Asymptotically in $m$, the saturation of Eq.~(\ref{CRLB}) is obtained for the
maximum likelihood estimator (MLE) \cite{KayBook, Lehmann,Pflanzagl}.
This is the value $\theta_{\mathrm{MLE}}(\bm{\mu})$
that maximizes the likelihood function $p(\bm{\mu}|\theta_0)$ (as a function of the parameter $\theta_0$)
for the observed measurement sequence $\bm{\mu}$,
\begin{equation} \label{MLE}
\theta_{\mathrm{MLE}}(\bm{\mu}) \equiv \mathrm{\arg \max_{\theta_0}} \{ p(\bm{\mu}|\theta_0) \},
\end{equation}
For a sufficiently large sample size $m$ (in the central limit), independently of the probability distribution $p(\bm{\mu}\vert \theta_0)$,
the MLE becomes normally distributed \cite{Varenna, KayBook, Lehmann,Pflanzagl}:
\begin{equation}
p(\theta_{\mathrm{MLE}}|\theta_{0})=\sqrt{\frac{mF\left(  \theta_{0}\right)  }{2\pi}
}e^{-\frac{mF\left(  \theta_{0}\right)  }{2}\left(  \theta_{0}-\theta
_{\mathrm{MLE}}\right) ^2 }\qquad (m\gg 1),  \label{gaus}
\end{equation}
with mean given by the true value $\theta_{0}$
and variance equal to the inverse of the Fisher information.

\begin{figure}[tb]
\centering
\includegraphics[width=1\textwidth]{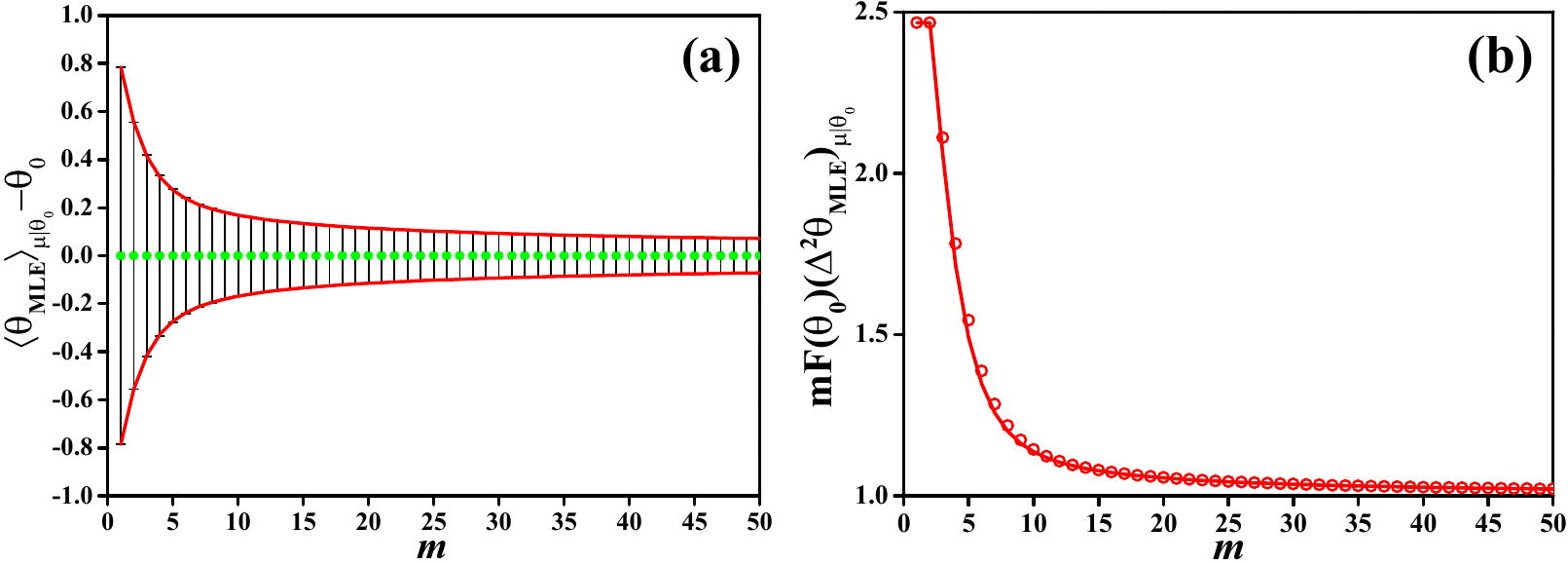}
\caption{(a) Bias $\langle \theta_{\rm MLE} \rangle_{\bm{\mu} \vert \theta_0} - \theta_0$ (green dots) as function of $m$ with error bars $(\Delta\theta_{\mathrm{MLE}})_{\bm{\mu} \vert \theta_0}$.
The red lines are $\pm \Delta \theta_{\rm CRLB} = \pm |d \langle \theta_{\rm MLE} \rangle_{\bm{\mu} \vert \theta_0} / d\theta_0|/\sqrt{mF(\theta_0)}$.
(b) Variance of the maximum likelihood estimator multiplied by the Fisher information, $ m F(\theta_0) (\Delta^2\theta_{\mathrm{MLE}})_{\bm{\mu} \vert \theta_0}$ (red circles),
as a function of the sample size $m$.
It is compared to the bias $(d \langle \theta_{\rm MLE} \rangle_{\bm{\mu} \vert \theta_0} / d\theta_0)^2$ (red line).
We recall that $\theta_0=\pi/4$ and $F(\theta_0)=4$ here.}
\label{fig01}
\end{figure}
In Fig.~\ref{fig01} we plot the results of a maximum likelihood analysis for the example considered in this manuscript.
In this case, the MLE is readily calculated and given by $\theta_{\mathrm{MLE}}(\bm{\mu}) =\frac{1}{2}\arccos (\frac{m_+-m_-}{m_++m_-} )$,
and the Fisher information is $F(\theta_0) = N^2$, independent of $\theta_0$.
In Fig.~\ref{fig01}(a) we plot the bias $\left\langle \theta_{\mathrm{MLE}} \right\rangle_{\bm{\mu} \vert \theta_0} - \theta_0$ (dots) as a function of $m$, for $\theta_0=\pi/4$.
Error bars are $\pm \Delta \theta_{\rm CRLB}$.
Notice that $\left\langle \theta_{\mathrm{MLE}} \right\rangle_{\bm{\mu} \vert \theta_0} = \theta_0$ for every $m$. 
This does not mean that the estimator is locally unbiased: indeed the derivative 
$d \left\langle \theta_{\mathrm{MLE}} \right\rangle_{\bm{\mu} \vert \theta_0} / d \theta_0$
[shown in panel (b)] is different from 1 for every value of $m$. 
We have $d \left\langle \theta_{\mathrm{MLE}} \right\rangle_{\bm{\mu} \vert \theta_0} / d \theta_0 \to 1$ asymptotically in $m$.
In Fig.~\ref{fig01}(b) we plot $m F(\theta_0)(\Delta^2\theta_{\mathrm{MLE}})_{\bm{\mu} \vert \theta_0}$ as a function of the number of independent measurements $m$ (red dots).
This quantity is compared to $m F(\theta_0) \Delta^{2}\theta_{\mathrm{CRLB}} = (d \left\langle \theta_{\mathrm{MLE}} \right\rangle_{\bm{\mu} \vert \theta_0} / d \theta_0)^2$ (red line).
With increasing sample size $m$, $(\Delta^2\theta_{\mathrm{MLE}})_{\bm{\mu} \vert \theta_0} \to 1/\big(m F(\theta_0)\big)$ corresponding to the CRLB for unbiased estimators.

\begin{figure}[tb]
\centering
\includegraphics[width=1\textwidth]{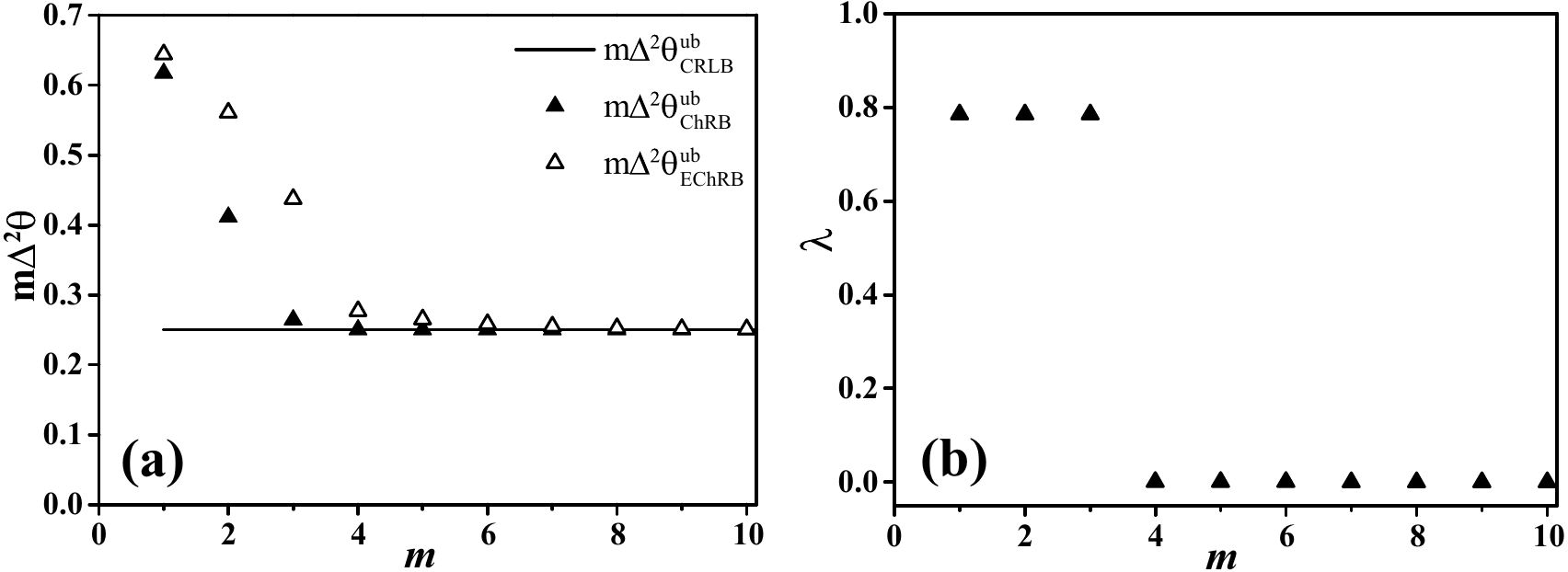}
\caption{(a) Comparison between unbiased frequentist bounds for the example considered in this manuscript, Eq.~(\ref{mlf}):
the CRLB $m\Delta^2 \theta^{\rm ub}_{\rm CRLB} = 1/F(\theta_0)$ (black line),
the Hammersley-Chapman-Robbins bound $m\Delta^2 \theta^{\rm ub}_{\rm ChRB}$ [Eq.~(\ref{CHRBub}), filled triangles]
and the extended Hammersley-Chapman-Robbins bound $m\Delta^2 \theta^{\rm ub}_{\rm EChRB}$ [Eq.~({\ref{EXCHRBub}}), empty triangles].
(b) Values of $\lambda$ achieving the supremum in Eq.~(\ref{CHRBub}), as a function of $m$.
}
\label{fig02}
\end{figure}

\subsubsection{Hammersley-Chapman-Robbins bound.}

The ChRB is obtained from Eq.~(\ref{BB})
by taking $n=2$, $a_1=1, a_2=-1$, $\theta_1=\theta_0+\lambda$, $\theta_2=\theta_0$, and can be written as \cite{Hammersley1950,Chapman1951}
\begin{equation}
\Delta^2 \theta_{\rm ChRB} =
\sup_{\lambda}  \frac{\left(\langle\theta_{\mathrm{est}}\rangle_{\boldsymbol{\mu}|\theta_0+\lambda}-\langle\theta_{\mathrm{est}}\rangle_{\boldsymbol{\mu}|\theta_0}\right)^{2}}{\sum_{\bm{\mu}}
\frac{p(\bm{\mu}|\theta_0+\lambda)^{2}}{p(\bm{\mu}|\theta_0)}-1}.  \label{CHRB}
\end{equation}
Clearly, restricting the number of parameters in the optimization in Eq.~(\ref{BB}) leads to a less strict bound.
We thus have $\Delta^2 \theta_{\rm BB} \geq \Delta^2 \theta_{\rm ChRB}$. For unbiased estimators, we obtain
\begin{equation}
\Delta^2 \theta_{\rm ChRB}^{\mathrm{ub}} =
\sup_{\lambda} \frac{\lambda^2}{\sum_{\bm{\mu}}
\frac{p(\bm{\mu}|\theta_0+\lambda)^{2}}{p(\bm{\mu}|\theta_0)}-1}. \label{CHRBub}
\end{equation}
Furthermore, the supremum over $\lambda$ on the right side of Eq.~(\ref{CHRB}) is always larger or equal to its limit $\lambda\to 0$:
\begin{eqnarray}
\sup_{\lambda} \frac{\left(\langle\theta_{\mathrm{est}}\rangle_{\boldsymbol{\mu}|\theta_0+\lambda}-\langle\theta_{\mathrm{est}}\rangle_{\boldsymbol{\mu}|\theta_0}\right)^{2}}{\sum_{\bm{\mu}}
\frac{p(\bm{\mu}|\theta_0+\lambda)^{2}}{p(\bm{\mu}|\theta_0)}-1}
&\geq
\lim_{\lambda\rightarrow0}\frac{\left(\langle\theta_{\mathrm{est}}\rangle_{\boldsymbol{\mu}|\theta_0+\lambda}-\langle\theta_{\mathrm{est}}\rangle_{\boldsymbol{\mu}|\theta_0}\right)^{2}}{\sum_{\bm{\mu}}
\frac{p(\bm{\mu}|\theta_0+\lambda)^{2}}{p(\bm{\mu}|\theta_0)}-1} \nonumber \\
&=
\frac{\left( \frac{d \langle \theta_{\mathrm{est}} \rangle_{\bm{\mu} \vert \theta_0} }{d\theta_0}\right)^2}{ m \sum_{\mu} \frac{1}{p(\mu|\theta_0)} \big(\frac{d p(\mu|\theta_0)}{d\theta_0}\big)^{2}},
\end{eqnarray}
provided that the derivatives on the right-hand side exist. We thus recover the CRLB as a limiting case of the ChRB.
The ChRB is always stricter than the CRLB and
we obtain the last inequality in the chain~(\ref{ineq}).
Notice that the CRLB requires
the probability distribution $p(\bm{\mu}|\theta_0)$ to be differentiable \cite{VanTreesBook2007} --
a condition that can be dropped for the ChRB and the more general BB.
Even if the distribution is regular, the above derivation shows that the ChRB, and more generally the BB, provide tighter error bounds than the CRLB.
With increasing $n$,
the BB becomes tighter and tighter and the CRLB represents the
the weakest bound in this hierarchy, which can be observed in Fig.~\ref{fig02}(a).
Next, we determine a stricter bound in this hierarchy.

\subsubsection{Extended Hammersley-Chapman-Robbins bound.}

We obtain the extended Hammersley-Chapman-Robbins bound (EChRB)
as a special case of Eq.~(\ref{BB}), by taking $n=3$, $a_1=1$, $a_2=A$, $a_3=-1$,
$\theta_1=\theta_0+\lambda_1$, $\theta_2=\theta_0+\lambda_2$, and $\theta_3=\theta_0$, giving
\begin{equation}
\fl \Delta^2 \theta_{\rm EChRB} =
\sup_{\lambda_1,\lambda_2,A}
\frac{\left( \langle\theta_{\mathrm{est}}\rangle_{\boldsymbol{\mu}|\theta_0+\lambda_1}+A\langle\theta_{\mathrm{est}}\rangle_{\boldsymbol{\mu}|\theta_0+\lambda_2} -
(1+A)\langle\theta_{\mathrm{est}}\rangle_{\boldsymbol{\mu}|\theta_0} \right)  ^2}{\sum_{\bm{\mu}}%
\frac{\left[  p(\bm{\mu}|\theta_0+\lambda_1)-p(\bm{\mu}|\theta_0)+A
p(\bm{\mu}|\theta_0+\lambda_2)\right]  ^2}{p(\bm{\mu}|\theta_0
)}},  \label{EXCHRB}
\end{equation}
where the supremum is taken over all possible $\lambda_1, \lambda_2 \in \mathbb{N}$ and $A \in  \mathbb{R}$. Since the ChRB is obtained from Eq.~(\ref{EXCHRB}) in
the specific case $A=0$, we have that $\Delta^2 \theta_{\rm EChRB} \geq \Delta^2 \theta_{\rm ChRB}$. For unbiased estimators, we obtain
\begin{equation}
\Delta^2 \theta_{\rm EChRB}^{\mathrm{ub}} =
\sup_{\lambda_1,\lambda_2,A}
\frac{\left(  \lambda_1+A \lambda_2\right)  ^2}{\sum_{\bm{\mu}}
\frac{\left[  p(\bm{\mu}|\theta_0+\lambda_1)-p(\bm{\mu}|\theta_0)+A
p(\bm{\mu}|\theta_0+\lambda_2)\right]  ^2}{p(\bm{\mu}|\theta_0
)}}.  \label{EXCHRBub}
\end{equation}
In Fig.~\ref{fig02}(a) we compare the different bounds for unbiased estimators and for the example considered in the manuscript:
the CRLB (black line), the ChRB (filled triangles) and the EChRB (empty triangles), satisfying the chain of inequalities (\ref{ineq}).
In Fig.~\ref{fig02}(b) we show the values of $\lambda$ for which the supremum is achieved in our case.


\section{Bayesian approach}\label{sec:Bayes}

The Bayesian approach makes use of the Bayes-Laplace theorem, which can be very simply stated and proved. 
The joint probability of two stochastic variables $\bm{\mu}$ and $\theta$ is symmetric:
$p(\bm{\mu},\theta)=p(\bm{\mu}|\theta)p(\theta)=p(\theta|\bm{\mu})p(\bm{\mu})=p(\theta,\bm{\mu})$, where $p(\theta)$ and $p(\bm{\mu})$ are the marginal distributions, obtained
by integrating the joint probability over one of the two variables, while $p(\bm{\mu}\vert \theta)$ and $p(\theta \vert \bm{\mu})$ are conditional distributions.

We recall that, in a phase inference problem, the set of measurement results $\bm{\mu}$ is generated by a fixed and unknown value $\theta_0$
according to the likelihood $p(\bm{\mu} \vert \theta_0)$. In the Bayesian approach to the estimation of $\theta_0$ one introduces a random variable $\theta$ and
uses the Bayes-Laplace theorem to define the conditional probability
\begin{equation}
\centering
p_{\mathrm{post}}(\theta|\bm{\mu}) = \frac{p(\bm{\mu}|\theta) p_{\mathrm{pri}}(\theta)}{p_{\rm mar}(\bm{\mu})}. \label{bayesianT}
\end{equation}
The posterior probability $p_{\mathrm{post}}(\theta|\bm{\mu})$
provides a degree of belief, or plausibility, that $\theta_0 = \theta$ (i.e., that $\theta$ is the true value of the phase), in the light of the measurement data $\bm{\mu}$ \cite{Sivia2006}.
In Eq.~(\ref{bayesianT}) the prior distribution $p_{\mathrm{pri}}(\theta)$ expresses the \textit{a priori} state of knowledge on $\theta$,
$p(\bm{\mu}|\theta)$ is the likelihood which is determined by the quantum mechanical measurement postulate, e.g., as in Eq.~(\ref{mlf}), and the marginal
probability $p_{\rm mar}(\bm{\mu}) = \int_a^b d \theta \, p(\theta, \bm{\mu})$ is obtained through the normalization for the posterior, where $a$ and $b$ are boundaries
of the phase domain. The posterior probability $p_{\mathrm{post}}(\theta|\bm{\mu})$
describes the current knowledge about the random variable $\theta$ based on
the available information, i.e., the measurement results $\bm{\mu}$.

\subsection{Noninformative prior}

In the Bayesian approach, the information on $\theta$ provided by the posterior probability always depends on the prior distribution $p_{\mathrm{pri}}(\theta)$.
It is possible to account for the available \textit{a priori} information on $\theta$ by choosing a prior distribution accordingly. However, if no \textit{a priori}
information is available, it is not obvious how to choose a ``noninformative'' prior \cite{CPRobert}. The flat prior $p_{\mathrm{pri}}(\theta)= \mathit{const}$ was first
introduced by Laplace to express the absence of information on $\theta$ \cite{CPRobert}. However, this prior would not be flat for other functions of $\theta$ and, in the
complete absence of \textit{a priori} information, it seems unreasonable that some information is available for different parametrizations of the problem. To see this recall
that a transformation of variables requires that $p_{\mathrm{pri}}(\varphi) = p_{\mathrm{pri}}(\theta)|df^{-1}(\varphi)/d\varphi|$ for any function $\varphi=f(\theta)$.
Hence, if $p_{\mathrm{pri}}(\theta)$ is flat, one obtains that $p_{\mathrm{pri}}(\varphi)=|df^{-1}(\varphi)/d\varphi|$ is, in general, not flat.

Notice that $p_{\mathrm{pri}}(\theta)\propto\sqrt{F(\theta)}$ -- called Jeffreys prior \cite{Jeffreys1946,Jeffreys1961} --
where $F(\theta)$ is the Fisher information~(\ref{fisher}), remains functionally invariant under changes of variable.
It is easy to check that $F(\varphi)=F(\theta)(d\theta/d\varphi)^2$ and, thus, $p_{\mathrm{pri}}(\varphi)\propto\sqrt{F(\varphi)}$ for arbitrary one-to-one
transformations $\varphi=f(\theta)$. Notice that, as in our case, the Fisher information $F(\theta)$ may actually be independent of $\theta$ (for a particular parametrization
of the problem). In this case, the invariance property does not imply that Jeffreys prior is flat for arbitrary reparametrizations $\varphi=f(\theta)$. Instead, it means that
for any $\varphi$ the prior will be proportional to $\sqrt{F(\varphi)}$, which, for $F(\theta)=\mathit{const}$ is given by $\sqrt{F(\varphi)}=|df^{-1}(\varphi)/d\varphi|$,
as expected by the transformation property of the flat prior.

\subsection{Posterior bounds}

From the posterior probability (\ref{bayesianT}), we can provide an estimate $\theta_{\rm BL}(\bm{\mu})$ of $\theta_0$.
This can be the maximum \textit{a posteriori},
$\theta_{\rm BL}(\bm{\mu}) = \mathrm{\arg \max_{\theta}} \, p_{\mathrm{post}}(\theta|\bm{\mu})$,
which coincides with the maximum likelihood Eq.~(\ref{MLE}) when the prior is flat, $ p_{\mathrm{pri}}(\theta) = {\it const}$,
or the mean of the distribution, $\theta_{\rm BL}(\bm{\mu}) =  \int_{a}^{b} d\theta \, \theta \, p_{\mathrm{post}}(\theta|\bm{\mu})$.

With the Bayesian approach it is possible to provide a confidence interval around the estimator, given an arbitrary measurement sequence $\bm{\mu}$,
even with a single measurement.
For instance the variance
\begin{equation} \label{BayVar}
\left(\Delta^{2}\theta_{\rm BL}(\bm{\mu})\right)_{\theta|\bm{\mu}} =
\int_{a}^{b} d\theta \, p_{\mathrm{post}}(\theta|\bm{\mu}) \big(  \theta - \theta_{\rm BL}(\bm{\mu}) \big)^{2},
\end{equation}
can be taken as a measure of fluctuation of our degree of belief around $\theta_{\rm BL}(\bm{\mu})$.
There is no such a concept in the frequentist paradigm.
The Bayesian posterior variance $\big(\Delta^{2}\theta_{\rm BL}(\bm{\mu})\big)_{\theta|\bm{\mu}}$
and the frequentist variance $( \Delta^{2}\theta_{\rm BL} )_{\bm{\mu} \vert \theta_0}$ have entirely different operational meanings.
Equation~(\ref{BayVar}) provides a degree of plausibility that $\theta_{\rm BL}(\bm{\mu})=\theta_0$, given the measurement results $\bm{\mu}$.
There no notion of bias in this case.
On the other hand, the quantity $( \Delta^{2}\theta_{\rm BL} )_{\bm{\mu} \vert \theta_0}$ measures the statistical fluctuations of $\theta_{\rm BL}(\bm{\mu})$
when repeating the sequence of $m$ measurements infinitely many times.

\subsubsection{Ghosh bound.}
In the following we derive a lower bound to Eq.~(\ref{BayVar}) first introduced by Ghosh \cite{Gosh}.
Using $ \int_{a}^{b}d\theta \, p_{\mathrm{post}}(\theta|\bm{\mu}) = 1$
we have
\begin{eqnarray}
\fl \int_{a}^{b}d\theta \, \left(  \theta-\theta_{\rm BL}(\bm{\mu})\right)  \frac{d
p_{\mathrm{post}}(\theta|\bm{\mu})}{d \theta} & =
\left.p_{\mathrm{post}}(\theta|\bm{\mu})\big(  \theta-\theta_{\rm BL}(\bm{\mu})\big)  \right|_{a}^{b}
-\int_{a}^{b} d\theta \, p_{\mathrm{post}}(\theta|\bm{\mu})
\nonumber\\
&  =f\left(  \bm{\mu},a,b\right)  -1,
\end{eqnarray}
where $f\left(  \bm{\mu},a,b\right)=b p_{\mathrm{post}}(b|\bm{\mu})-a p_{\mathrm{post}}(a|\bm{\mu})-
\theta_{\rm BL}(\bm{\mu})(p_{\mathrm{post}}(b|\bm{\mu})-p_{\mathrm{post}}(a|\bm{\mu}))$
depends on the value of the posterior distribution calculated at the boundaries.
If $p_{\mathrm{pri}}(a)=p_{\mathrm{pri}}(b)=0$, we have $f\left(\bm{\mu},a,b\right)=0$.
In analogy with the derivation of the (frequenstist) CRLB, we exploit the Cauchy-Schwarz inequality,
\begin{equation*}
\fl \left(\int_{a}^{b} d\theta \left(
\frac{d p_{\mathrm{post}}(\theta|\bm{\mu})}{d \theta}\right)  ^{2}\frac{1}{p_{\mathrm{post}}(\theta|\bm{\mu})}\right)
\left(\int_{a}^{b} d \theta \, p_{\mathrm{post}}(\theta|\bm{\mu})\left(  \theta-\theta_{\rm BL}(\bm{\mu})\right)  ^{2} \right)
\geq(f\left(  \bm{\mu},a,b\right)  -1)^{2},
\end{equation*}
leading to $(\Delta^{2}\theta_{\rm BL}(\bm{\mu}))_{\theta|\bm{\mu}} \geq \Delta^2 \theta_{\rm GB}(\bm{\mu})$, where~\cite{Gosh}
\begin{equation}
\Delta^2 \theta_{\rm GB}(\bm{\mu}) =
\frac{(f\left(
\bm{\mu},a,b\right)  -1)^{2}}{\int_{a}^{b} d \theta \, \frac{1}{p_{\mathrm{post}}(\theta|\bm{\mu})} \left(  \frac{d p_{\mathrm{post}}(\theta|\bm{\mu}
)}{d \theta}\right)  ^{2} }. \label{pcrlb0}
\end{equation}
The above bound is a function of the specific measurement sequence $\bm{\mu}$ and depends on $\int_{a}^{b} d \theta \, \frac{1}{p_{\mathrm{post}}(\theta|\bm{\mu})}
\big(  \frac{d p_{\mathrm{post}}(\theta|\bm{\mu})}{d \theta} \big)^{2}$ that we can identify as a ``Fisher information of the posterior distribution''. 
The Ghosh bound is saturated
if and only if
\begin{equation} \label{GBsaturation}
\theta-\theta_{\rm BL}(\bm{\mu}) = \lambda_{\bm{\mu}} \frac{d \log p(\theta \vert \bm{\mu})}{d \theta},
\end{equation}
where $\lambda_{\bm{\mu}}$ does not depend on $\theta$ while it may depend on $\bm{\mu}$.

\subsection{Average Posterior bounds}

While Eq.~(\ref{BayVar}) depends on the specific $\bm{\mu}$,
it is natural to consider its average over all possible measurement sequences at fixed $\theta_0$ and $m$, weighted by the likelihood $p(\bm{\mu} \vert \theta_0)$:
\begin{equation} \label{Bayvar}
\fl \left( \Delta^{2}\theta_{\rm BL} \right)_{\bm{\mu},\theta \vert \theta_0} =
\sum_{\bm{\mu}} \big( \Delta^2 \theta_{\rm BL}(\bm{\mu}) \big)_{\theta \vert \bm{\mu}} \, p(\bm{\mu} \vert \theta_0)
= \sum_{\bm{\mu}} \int_{a}^{b} d \theta \,
p(\theta, \bm{\mu} |\theta_0)\big(  \theta-\theta_{\rm BL}(\bm{\mu})\big)  ^{2},
\end{equation}
that we indicate as average Bayesian posterior variance, where $p(\theta,\bm{\mu}|\theta_0)=p_{\rm post}(\theta|\bm{\mu})p(\bm{\mu}|\theta_0)$.

We would be tempted to compare the average posterior sensitivity $( \Delta^{2}\theta_{\rm BL} )_{\bm{\mu},\theta\vert \theta_0}$
to the frequentist Cram\'er-Rao bound $\Delta^2 \theta_{\rm CRLB}$.
However, because of the different operational meaning between the frequentist and the Bayesian paradigms,
there is no reason for Eq.~(\ref{Bayvar}) to fulfill the Cram\'er-Rao bound: indeed it does not, as we show below.

\subsubsection{Likelihood-averaged Ghosh bound.}
A lower bound to Eq.~(\ref{Bayvar}) is obtained by averaging the Ghosh bound Eq.~(\ref{pcrlb0}) over the likelihood function.
We have $( \Delta^2 \theta_{\rm BL} )_{\bm{\mu},\theta \vert \theta_0 } \geq \Delta^2 \theta_{\rm aGB}$, where \cite{Varenna}
\begin{equation} \label{pcrlb}
 \Delta^2 \theta_{\rm aGB} =
\sum_{\bm{\mu}} \frac{(f\left(  \bm{\mu},a,b\right)
-1)^{2}}{\int_{a}^{b} d \theta \, \frac{1}{p_{\mathrm{post}}(\theta|\bm{\mu})}
\big(  \frac{\partial p_{\mathrm{post}}(\theta|\bm{\mu})}{\partial\theta} \big)^{2}} \, p(\bm{\mu}|\theta_{0}).
\end{equation}
This likelihood-averaged Ghosh bound is independent of $\bm{\mu}$ because of the statistical average.

\subsection{Numerical comparison of Bayesian and frequentist phase estimation}

\begin{figure}[tbp]
\centering
\includegraphics[width=1\textwidth]{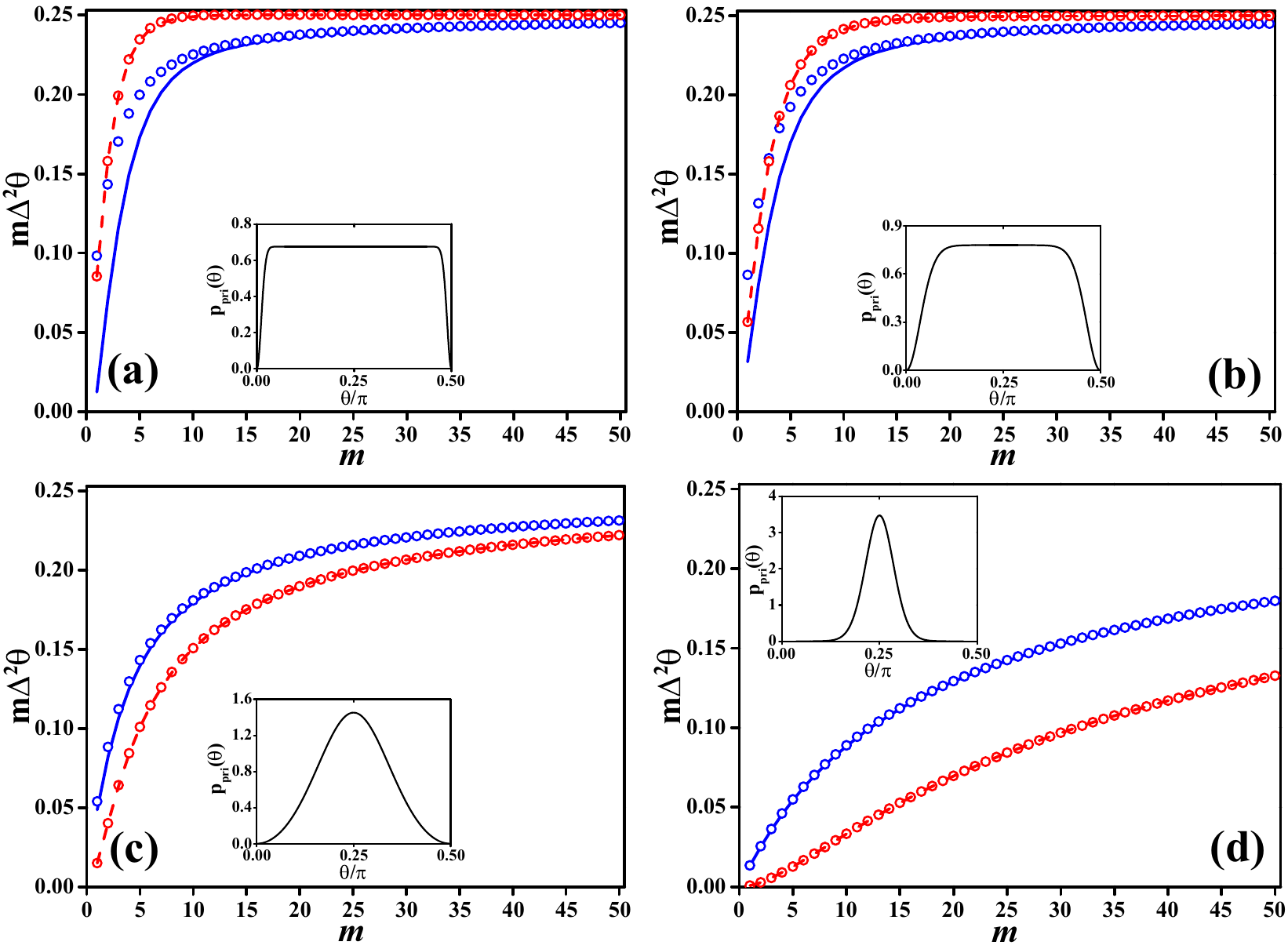}
 \caption{Phase estimation variance as a function of the sample size for Bayesian and frequentist data analysis.
 Red circles (frequentist) are $m (\Delta^2 \theta_{\rm BL})_{\bm{\mu} \vert \theta_0}$, the red dashed line is the CRLB $m\Delta^2 \theta_{\rm CRLB}$, Eq.~(\ref{CRLB}).
 Blue circles (Bayesian) are $m (\Delta^2 \theta_{\rm BL})_{\bm{\mu},\theta \vert \theta_0}$, the blue solid line is the likelihood-averaged Ghosh bound  $m\Delta^2 \theta_{\rm aGB}$, Eq.~(\ref{pcrlb}).
The insets show the prior distribution $p_{\rm pri}(\theta)$.}
\label{fig03}
\end{figure}

In the numerical calculations shown in Fig.~\ref{fig03} we consider a Bayesian estimator given by
$ \theta_{\rm BL} (\bm{\mu}) =   \int_{a}^{b} d\theta \, \theta \, p_{\mathrm{post}}(\theta|\bm{\mu})$
with prior distributions 
\begin{equation}\label{exampleprior}
p_{\rm pri}(\theta) = \frac{2}{\pi} \frac{e^{\alpha\sin(2 \theta)^2}-1}{e^{\alpha/2}  I_0(\alpha/2)-1}.
\end{equation}
Equation (\ref{exampleprior}) is normalized to one for $\theta\in[0,\frac{\pi}{2}]$, 
where $I_0(\alpha)$ is the Bessel function.
The more negative is $\alpha$, the more $p_{\rm pri}(\theta)$ broadens in $[0, \pi/2]$. 
In particular, in the limit $\alpha\rightarrow -\infty$ the prior approaches the flat distribution,
which in our case coincides with Jeffreys prior since the Fisher information is independent of $\theta$.
In the limit $\alpha=0$, the prior is given by $\lim_{\alpha\to 0} p_{\mathrm{pri}}(\theta)=4 \sin(2 \theta)^2/\pi$. 
For positive values of $\alpha$, the larger $\alpha$, the more peaked is $p_{\rm pri}(\theta)$ around $\theta_0 = \pi/4$. 
In particular $p_{\rm pri}(\theta) \approx e^{-4 \alpha (\theta - \pi/4)^2}/\sqrt{\pi/4 \alpha}$ for $\alpha \gg 1$.
In the inset of the different panels of Fig.~\ref{fig03} we plot $p_{\rm pri}(\theta)$ for 
$\alpha=-100$ [panel (a)], $\alpha=-10$ (b), $\alpha=1$ (c) and $\alpha=10$ (d).

In Fig.~\ref{fig03} we plot, as a function of $m$,
the posterior variance $(\Delta^2 \theta_{\rm BL})_{\bm{\mu},\theta\vert \theta_0}$ (blue circles) that, as expected, is always larger than the
likelihood-averaged Ghosh bound Eq.~(\ref{pcrlb}) (solid blue lines).
For comparison, we also plot the frequentist variance $( \Delta^{2}\theta_{\rm BL} )_{\bm{\mu} \vert \theta_0}  = \sum_{\bm{\mu}} \big( \theta_{\rm BL}(\bm{\mu}) -
\langle \theta_{\rm BL} \rangle_{\bm{\mu}|\theta_0} \big)^2 p(\bm{\mu} \vert \theta_0)$
(red dots) around the mean value $\langle \theta_{\rm BL} \rangle_{\bm{\mu}|\theta_0} = \sum_{\bm{\mu}} \theta_{\rm BL}(\bm{\mu}) p(\bm{\mu} \vert \theta_0)$ of the estimator.
This quantity obeys the Cram\'er-Rao theorem $\left( \Delta^{2}\theta_{\rm BL} \right)_{\bm{\mu} \vert \theta_0} \geq  \Delta^2 \theta_{\rm CRLB}$ and the more general chain of inequalities (\ref{ineq}).
This is confirmed in the figure where we show $\Delta^2 \theta_{\rm CRLB} =
|d \langle \theta_{\rm BL} \rangle_{\bm{\mu}|\theta_0} / d\theta_0|^2/\big(m F(\theta_0)\big)$
(red line).
Notice that, when the prior narrows around $\theta_0$, the variance $\left( \Delta^{2}\theta_{\rm BL} \right)_{\bm{\mu} \vert \theta_0}$ decreases but,
at the same time, the estimator becomes more and more biased, i.e. $|d \langle \theta_{\rm BL} \rangle_{\bm{\mu}|\theta_0} / d\theta_0|$ decreases as well.

Interestingly, in Fig.~\ref{fig03} we clearly see that the Bayesian posterior variance $(\Delta^2 \theta_{\rm BL})_{\bm{\mu},\theta\vert \theta_0}$ and the
likelihood-averaged Ghosh bound may stay in some cases below the (frequentist) $\Delta^2 \theta_{\rm CRLB}$, even if the prior is flat in the full phase interval $[0, \pi/2]$.
The discrepancy with the CRLB is remarkable and can be quite large for small values of $m$.
Still, there is no contradiction since $(\Delta^2 \theta_{\rm BL})_{\bm{\mu},\theta\vert \theta_0}$ and $\left( \Delta^{2}\theta_{\rm BL} \right)_{\bm{\mu} \vert \theta_0}$
have different operational meanings and interpretations. They both respect their corresponding sensitivity bounds.

Asymptotically in the number of measurements $m$, the Ghosh bound as well as its likelihood average
converge to the Cram\'er-Rao bound.
Indeed, it is well known that in this limit the posterior probability becomes a
Gaussian centered at the true value of the phase shift and with variance given by the inverse of the Fisher information,
\begin{equation} \label{LBVM}
p_{\mathrm{post}}(\theta|\bm{\mu}) = \sqrt{\frac{m F(\theta_0)}{2 \pi}} e^{-\frac{m F(\theta_0)}{2} (\theta-\theta_0)^2}, \qquad (m \gg 1)
\end{equation}
a results known as Laplace-Bernstein-von Mises theorem \cite{Varenna, Lehmann, LeCam}.
By replacing Eq.~(\ref{LBVM}) into Eq.~(\ref{pcrlb0}), we recover a posterior variance given by $1/\big(mF(\theta_0)\big)$.


\section{Bounds for random parameters}\label{sec:rand}

In this section we derive bounds of phase sensitivity
obtained when $\theta_0$ is a random variable distributed according to $p(\theta_0)$.
Operationally, this corresponds to the situation where $\theta_0$ remains fixed (but unknown) when collecting a single sequence of $m$ measurements $\bm{\mu}$.
In between measurement sequences, $\theta_0$ fluctuates according to $p(\theta_0)$.

\subsection{Frequentist risk functions for random parameters}
Let us first consider the frequentist estimation of a fluctuating parameter $\theta_0$ with the estimator $\theta_{\mathrm{est}}$.
The mean sensitivity obtained by averaging $(\Delta^2 \theta_{\rm est})_{\bm{\mu}\vert \theta_0}$, Eq.~(\ref{MSE}), over $p(\theta_0)$ is
\begin{eqnarray}
(\Delta^2 \theta_{\rm est})_{\bm{\mu}, \theta_0} &=& \int_{a}^b d \theta_0(\Delta^2 \theta_{\rm est})_{\bm{\mu}\vert \theta_0} p(\theta_0)  \nonumber \\
&=&
\sum_{\bm{\mu}} \int_{a}^b d \theta_0 \, p(\bm{\mu}\vert \theta_0) p(\theta_0)
\big( \langle \theta_{\rm est}  \rangle_{\bm{\mu}|\theta_0} - \theta_{\rm est}(\bm{\mu})  \big)^2  \nonumber \\
&=&
 \sum_{\bm{\mu}} \int_{a}^b d \theta_0 \, p(\bm{\mu}, \theta_0)
\big( \langle \theta_{\rm est}  \rangle_{\bm{\mu}|\theta_0} - \theta_{\rm est}(\bm{\mu})  \big)^2, \label{avevar}
\end{eqnarray}
where $\bm{\mu}$ and $\theta_0$ are both random variables and we have used $p(\bm{\mu} \vert \theta_0) p(\theta_0) = p(\bm{\mu}, \theta_0)$.

An averaged risk function for the efficiency of the estimator is given by averaging the mean square error~(\ref{MSE}) over $p(\theta_0)$, leading to
\begin{equation} \label{avgMSE}
\fl \mathrm{MSE}(\theta_{\mathrm{est}})_{\bm{\mu},\theta_0}=\int d\theta_0\mathrm{MSE}(\theta_{\mathrm{est}})_{\bm{\mu}|\theta_0}p(\theta_0)=\int d\theta_0
\sum_{\bm{\mu}} \big( \theta_{\mathrm{est}}(\bm{\mu}) - \theta_0 \big)^2 p(\bm{\mu},\theta_0).
\end{equation}
In analogy to Eq.~(\ref{varMSEbias}), we can write
\begin{equation} \label{avgvarMSEbias}
\mathrm{MSE}(\theta_{\mathrm{est}})_{\bm{\mu},\theta_0}=\big( \Delta^2 \theta_{\rm est} \big)_{\bm{\mu} , \theta_0} +
\int d\theta_0\left(\langle\theta_{\mathrm{est}}\rangle_{\boldsymbol{\mu}|\theta_0}-\theta_0\right)^2p(\theta_0).
\end{equation}

In the following, we derive lower bounds for both $(\Delta^2 \theta_{\rm est})_{\bm{\mu}, \theta_0}$ and $\mathrm{MSE}(\theta_{\mathrm{est}})_{\bm{\mu},\theta_0}$.
Notice that bounds on $(\Delta^2 \theta_{\rm est})_{\bm{\mu}, \theta_0}$ hold also for $\mathrm{MSE}(\theta_{\mathrm{est}})_{\bm{\mu},\theta_0}$ due to
$\mathrm{MSE}(\theta_{\mathrm{est}})_{\bm{\mu},\theta_0} \geq (\Delta^2 \theta_{\rm est})_{\bm{\mu}, \theta_0}$.
Nevertheless, bounds on the average the mean square error are widely used (and are often called Bayesian bounds \cite{VanTreesBOOK1968}) since they
can be expressed independently of the bias.

\subsection{Bounds on the mean square error}
We first consider bounds on $\mathrm{MSE}(\theta_{\mathrm{est}})_{\bm{\mu},\theta_0}$, Eq.~(\ref{avgMSE}), for arbitrary estimators.

\subsubsection{Van Trees bound.}\label{sec:vT}
It is possible to derive a general lower bound on the mean square error~(\ref{avgMSE}) based on the following assumptions:
\begin{enumerate}
\item $\frac{\partial p(\bm{\mu},\theta_0)}{\partial\theta_0}$ and $\frac{\partial^{2} p(\bm{\mu},\theta_0)}{\partial\theta_0^{2}}$ are
absolutely integrable with respect to $\bm{\mu}$ and $\theta_0$;
\item $p\left(  a\right)  \xi(a) - p\left(  b\right)  \xi(b)=0$, where
$\xi(\theta_0)=\sum_{\bm{\mu}}\left(\theta_{\mathrm{est}}(\bm{\mu})- \theta_0 \right)  p(\bm{\mu}|\theta_0)$.
\end{enumerate}
Multiplying $\xi(\theta_0)$ by $p(\theta_0)$ and differentiating with respect to $\theta_0$, we have
\begin{equation*}
\frac{\partial p(  \theta_0)  \xi(\theta_0)}{\partial\theta_0}=
\sum_{\bm{\mu}}
\left(\theta_{\mathrm{est}}(\bm{\mu})- \theta_0 \right)  \frac{\partial p(\bm{\mu}, \theta_0)}{\partial \theta_0}
-
p(\theta_0). \nonumber
\end{equation*}
Integrating over $\theta_0$ in the range of $[a,b]$ and considering the above properties, we find
\begin{equation}
\sum_{\bm{\mu}} {\int_{a}^{b}} d \theta_0
\left(\theta_{\mathrm{BL}}(\bm{\mu})- \theta_0 \right)  \frac{\partial p(\bm{\mu}, \theta_0)}{\partial \theta_0} = 1.
\end{equation}
Finally, using the Cauchy-Schwarz inequality, we arrive at $\mathrm{MSE}(\theta_{\mathrm{est}})_{\bm{\mu},\theta_0}\geq  \Delta^2 \theta_{\rm VTB}$, where
\begin{equation} \label{VanTreesB}
\Delta^2 \theta_{\rm VTB} =
\frac{1}
 {\sum_{\bm{\mu}} \int_{a}^{b} d \theta_0 \frac{1}{p(\bm{\mu},\theta_0)} \big( \frac{\partial p(\bm{\mu},\theta_0)}{\partial\theta_0} \big)^{2}}
\end{equation}
is generally indicated as Van Trees bound~\cite{VanTreesBook2007,VanTreesBOOK1968,ShutzenbergerBAMS1957}.
The equality holds if and only if
\begin{equation}
\theta_{\rm est}(\bm{\mu}) - \theta_0 = \lambda \frac{d \log p(\bm{\mu}, \theta_0)}{d \theta_0},
\end{equation}
where $\lambda$ does not depend on $\theta_0$ and $\bm{\mu}$.
It is easy to show that
\begin{equation} \label{FishVTB}
\fl \sum_{\bm{\mu}} {\int_{a}^{b}} d \theta_0
 \frac{1}{p(\bm{\mu},\theta_0)}
\left(\frac{\partial p(\bm{\mu},\theta_0)}{\partial\theta_0}\right)^{2}
= m \int_a^b d\theta_0 \, p(\theta_0) F(\theta_0)
+{\int_{a}^{b}} d \theta_0 \frac{1}{p(\theta_0)} \left(  \frac{\partial p(\theta_0)}{\partial\theta_0}\right)^{2},
\end{equation}
where the first term is the Fisher information $F(\theta_0)$, defined by Eq.~(\ref{fisher}), averaged over $p(\theta_0)$, and
the second term can be interpreted as a Fisher information of the prior \cite{VanTreesBook2007}.
Asymptotically in the number of measurements $m$ and for regular distributions $p(\theta_0)$, the first term in Eq.~(\ref{FishVTB}) dominates over the second one.

\subsubsection{Ziv-Zakai bound.}
A further bound on $\mathrm{MSE}(\theta_{\mathrm{est}})_{\bm{\mu},\theta_0}$ can be derived by mapping the phase estimation problem to a continuous series of binary hypothesis
testing problems.
A detailed derivation of the Ziv-Zakai bound \cite{VanTreesBook2007,ZivZakai,Bell1997} is provided in the Appendix B. The final result reads
$\mathrm{MSE}(\theta_{\mathrm{est}})_{\bm{\mu},\theta_0} \geq \Delta^2\theta_{\rm ZZB}$, where
\begin{equation} \label{zivzakai}
\Delta^2\theta_{\rm ZZB}= \frac{1}{2}
\int dh \, h
\int
d\theta_0\left(  p\left(  \theta_0\right)  +p\left(  \theta_0+h\right)
\right) P_{\min}\left(  \theta_0,\theta_0+h\right),
\end{equation}
and
\begin{equation}
\fl P_{\mathrm{min}}\left(
\theta_0,\theta_0+h\right)  =\frac{1}{2}\left(  1-  \sum_{\bm{\mu}} \left\vert \frac{p\left(  \theta_0\right)  p\left(
\bm{\mu}|\theta_0\right)  }{p\left(  \theta_0\right)  +p\left(
\theta_0+h\right)  }-\frac{p\left(  \theta_0+h\right)  p\left(
\bm{\mu}|\theta_0+h\right)  }{p\left(  \theta_0\right)  +p\left(
\theta_0+h\right)  }\right\vert \right) \label{zivzakai1}
\end{equation}
is the minimum error probability of the binary hypothesis testing problem.
This bound has been adopted for quantum phase estimation in Ref.~\cite{Tsang2012}. To this end, the probability $P_{\mathrm{min}}(\theta_0,\theta_0+h)$ can be maximized over
all possible quantum measurements, which leads to the trace distance \cite{HelstromBOOK1976}. As the optimal measurement may depend on $\theta_0$ and $h$, the bound~(\ref{zivzakai})
which involves integration over all values of $\theta_0$ and $h$, is usually not saturable. We remark that the trace distance also defines a saturable frequentist bound for a
different risk function than the variance \cite{GS17}.

\subsection{Bounds on the average estimator variance}
We now consider bounds on $(\Delta^2 \theta_{\rm est})_{\bm{\mu}, \theta_0}$, Eq.~(\ref{avevar}), for arbitrary estimators.

\subsubsection{Average CRLB.}
Taking the average over $p(\theta_0)$ of Eq.~(\ref{ineq}), we obtain a chain of bounds for $(\Delta^2 \theta_{\rm est})_{\bm{\mu}, \theta_0}$.
In particular, in its simplest form we have $(\Delta^2 \theta_{\rm est})_{\bm{\mu}, \theta_0} \geq \Delta^2 \theta_{\rm aCRLB}$, where
\begin{equation} \label{aCRLB}
\Delta^2 \theta_{\rm aCRLB} = \int_a^b d \theta_0  \frac{ \left( \frac{d \langle \theta_{\mathrm{est}} \rangle_{\bm{\mu}|\theta_0} }{d\theta_0} \right)^2}{m F(\theta_0)} p(\theta_0),
\end{equation}
is the average CRLB.

\subsubsection{Van Trees bound for the average estimator variance.}
We can derive a general lower bound for the variance~(\ref{avevar}) by following the derivation of the Van Trees bound, which was discussed in Sec.~\ref{sec:vT}. In contrast to
the standard Van Trees bound for the mean square error, here the bias enters explicitly.
Defining $\xi(\theta_0)=\sum_{\bm{\mu}}\left(\theta_{\mathrm{est}}(\bm{\mu})- \langle \theta_{\rm est} \rangle_{\bm{\mu}|\theta_0}\right)  p(\bm{\mu}|\theta_0)$
and assuming the same requirements as in the derivation of the Van Trees bound for the MSE, we arrive at
\begin{equation*}
\sum_{\bm{\mu}} {\int_{a}^{b}} d \theta_0
\big(\theta_{\mathrm{est}}(\bm{\mu})- \langle \theta_{\rm est} \rangle_{\bm{\mu}|\theta_0} \big)  \frac{\partial p(\bm{\mu}, \theta_0)}{\partial \theta_0}
=
\int_{a}^{b} d \theta_0 \frac{d \langle \theta_{\rm est} \rangle_{\bm{\mu}|\theta_0}}{d \theta_0} p(\theta_0), \nonumber
\end{equation*}
Finally, a Cauchy-Schwarz inequality gives $(\Delta^2 \theta_{\rm est})_{\bm{\mu},\theta_0} \geq \Delta^2 \theta_{\rm fVTB}$, where
\begin{equation} \label{VanTrees}
\Delta^2 \theta_{\rm fVTB} =
 \frac{ \big( {\int_{a}^{b}} d \theta_0\frac{d \langle \theta_{\rm est} \rangle_{\bm{\mu}|\theta_0}}{d \theta_0} p(\theta_0) \big)^2}
 {\sum_{\bm{\mu}} \int_{a}^{b} d \theta_0 \frac{1}{p(\bm{\mu},\theta_0)} \big( \frac{\partial p(\bm{\mu},\theta_0)}{\partial\theta_0} \big)^{2}},
\end{equation}
with equality if and only if
\begin{equation}
\theta_{\rm est}(\bm{\mu}) - \langle \theta_{\rm est} \rangle_{\bm{\mu}|\theta_0} = \lambda \frac{d \log p(\bm{\mu}, \theta_0)}{d \theta_0},
\end{equation}
where $\lambda$ is independent of $\theta_0$ and $\bm{\mu}$.

We can compare Eq.~(\ref{VanTrees}) with the average CRLB Eq.~(\ref{aCRLB}).
We find
\begin{equation*}
\fl \int_a^b d \theta_0  \frac{ \big( \frac{d \langle \theta_{\mathrm{est}} \rangle_{\bm{\mu}|\theta_0} }{d\theta_0} \big)^2}{m F(\theta_0)} p(\theta_0)
 \geq   \frac{ \big(  {\int_{a}^{b}} d \theta_0\frac{d \langle \theta_{\rm est} \rangle_{\bm{\mu}|\theta_0}}{d \theta_0} p(\theta_0) \big)^2}
{m \int_{a}^{b} d \theta_0 p( \theta_0)  F(\theta_0)} \geq
\frac{ \big(  {\int_{a}^{b}} d \theta_0  \big| \frac{d \langle \theta_{\rm est} \rangle_{\bm{\mu}|\theta_0}}{d \theta_0} \big| p(\theta_0) \big)^2}
{\sum_{\bm{\mu}} \int_{a}^{b} d \theta_0 \frac{1}{p(\bm{\mu},\theta_0)} \big( \frac{\partial p(\bm{\mu},\theta_0)}{\partial\theta_0} \big)^{2}}, \nonumber
\end{equation*}
where in the first step we use Jensen's inequality, and the second step follows from Eq.~(\ref{FishVTB}) which implies
$m \int_{a}^{b} d \theta_0 p( \theta_0)  F(\theta_0) \leq
\sum_{\bm{\mu}} \int_{a}^{b} d \theta_0 \frac{1}{p(\bm{\mu},\theta_0)} \big( \frac{\partial p(\bm{\mu},\theta_0)}{\partial\theta_0} \big)^{2}$
since $\int_{a}^{b} d \theta_0 \frac{1}{p(\theta_0)} \big( \frac{d p(\theta_0)}{d \theta_0} \big)^{2} \geq 0$.
Finally, $\big(  {\int_{a}^{b}} d \theta_0  \big| \frac{d \langle \theta_{\rm est} \rangle_{\bm{\mu}|\theta_0}}{d \theta_0} \big| p(\theta_0) \big)^2 \geq
\big(  {\int_{a}^{b}} d \theta_0  \frac{d \langle \theta_{\rm est} \rangle_{\bm{\mu}|\theta_0}}{d \theta_0} p(\theta_0) \big)^2$ due to the triangle inequality.
We thus arrive at
\begin{equation} \label{ineq2}
(\Delta^2 \theta_{\rm est})_{\bm{\mu}, \theta_0} \geq \Delta^2 \theta_{\rm aCRLB} \geq \Delta^2 \theta_{\rm fVTB},
\end{equation}
that is valid for generic biased estimators.

\subsection{Bayesian framework for random parameters}
The Bayesian posterior variance, $(\Delta^2 \theta_{\rm BL})_{\bm{\mu},\theta \vert \theta_0}$, Eq.~(\ref{Bayvar}), averaged over $p(\theta_0)$ is
\begin{eqnarray} \label{Bave}
(\Delta^2 \theta_{\rm BL})_{\bm{\mu},\theta,\theta_0} &=& \int_{a}^b d \theta_0 (\Delta^2 \theta_{\rm BL})_{\bm{\mu},\theta \vert \theta_0} \, p(\theta_0)  \nonumber \\
&=& \sum_{\bm{\mu}} \int_{a}^b d \theta  \int_{a}^b d \theta_0 \,  p_{\rm post}(\theta \vert \bm{\mu}) p(\bm{\mu} \vert \theta_0) p(\theta_0) \big( \theta  - \theta_{\rm BL}(\bm{\mu}) \big)^2 \nonumber \\
&=& \sum_{\bm{\mu}} \int_{a}^b d \theta \, p_{\rm post}(\theta \vert \bm{\mu}) p(\bm{\mu}) \big( \theta  - \theta_{\rm BL}(\bm{\mu}) \big)^2,
\end{eqnarray}
where $p(\bm{\mu}) =\int_{a}^b d \theta_0 \,  p(\bm{\mu} \vert \theta_0) p(\theta_0)$ is the average probability to observe $\bm{\mu}$ taking into account fluctuations of $\theta_0$.

A bound on Eq.~(\ref{Bave}) can be obtained by averaging Eq.~(\ref{pcrlb}) over $p(\theta_0)$, or,
equivalently, averaging the Ghosh bound, Eq.~(\ref{pcrlb0}), over $p(\bm{\mu})$.
We obtain the average Ghosh bound for random parameters $\theta_0$,
$( \Delta^{2}\theta_{\rm BL} )_{\bm{\mu},\theta,\theta_0} \geq \Delta^2 \theta_{\rm aGBr}$, where
\begin{eqnarray}
\Delta^2 \theta_{\rm aGBr} &= &
\int_{a}^{b} d \theta_0 \sum_{\bm{\mu}} \frac{(f\left(  \bm{\mu},a,b\right)  -1)^{2}}{\int_{a}^{b} d \theta \frac{1}{p_{\mathrm{post}}(\theta\vert \bm{\mu})}
\left(\frac{d p_{\mathrm{post}}(\theta|\bm{\mu})}{d \theta}\right)^{2}} p(\bm{\mu} \vert \theta_0) p(\theta_0)  \nonumber \\
&=&
\sum_{\bm{\mu}}  \frac{(f\left(  \bm{\mu},a,b\right)  -1)^{2}}
{\int_{a}^{b} d\theta \, \frac{1}{p_{\mathrm{post}}(\theta|\bm{\mu})} \left(\frac{d p_{\mathrm{post}}(\theta|\bm{\mu})}{d \theta}\right)  ^{2}} p(\bm{\mu}). \label{avpcrlb}
\end{eqnarray}
The bound holds for any prior $p_{\rm pri}(\theta)$ and
is saturated if and only if, for every value of $\bm{\mu}$, there exists a $\lambda_{\bm{\mu}}$ such that Eq.~(\ref{GBsaturation}) holds.

\subsubsection{Bayesian bounds.}
In Eq.~(\ref{Bave}), the prior used to define the posterior $p_{\rm post}(\theta \vert \bm{\mu})$ via the Bayes-Laplace theorem is arbitrary.
In general, such a prior $p_{\rm pri}(\theta)$ is different from the statistical distribution of $\theta_0$, which can be unknown.
If $p(\theta_0)$ is known, then one can use it as a prior in the Bayesian posterior probability, i.e., $p_{\rm pri}(\theta)= p(\theta_0)$.
In this specific case, we have $ p_{\rm mar}(\bm{\mu}) = p(\bm{\mu})$, and thus
$p_{\rm post}(\theta\vert \bm{\mu})p(\bm{\mu})  = p_{\rm post}(\theta\vert \bm{\mu})p_{\rm mar}(\bm{\mu}) = p(\bm{\mu},\theta)$.
In other words, for this specific choice of prior, the physical joint probability $p(\bm{\mu},\theta_0)$ of random variables $\theta_0$ and $\bm{\mu}$
coincides with the Bayesian $p(\bm{\mu},\theta)$.
Equation~(\ref{Bave}) thus simplifies to
\begin{equation} \label{Bayvarave}
(\Delta^2 \theta_{\rm BL})_{\bm{\mu},\theta}  = \sum_{\bm{\mu}}  \int_{a}^b d \theta  \, p(\bm{\mu},\theta) \big( \theta  - \theta_{\rm BL}(\bm{\mu}) \big)^2.
\end{equation}
Notice that this expression is mathematically equivalent to the frequentist average mean square error~(\ref{avgMSE}) if we replace $\theta$ with $\theta_0$ and
$\theta_{\rm BL}(\bm{\mu})$ with $\theta_{\mathrm{est}}(\bm{\mu})$.
This means that precision bounds for Eq.~(\ref{avgMSE}), e.g., the Van Trees and Ziv-Zakai bounds can also be applied to Eq.~(\ref{Bayvarave}).
These bounds are indeed often referred to as ``Bayesian bounds'', see Ref.~\cite{VanTreesBook2007}.

\begin{figure}[tb]
\centering
\includegraphics[width=\textwidth]{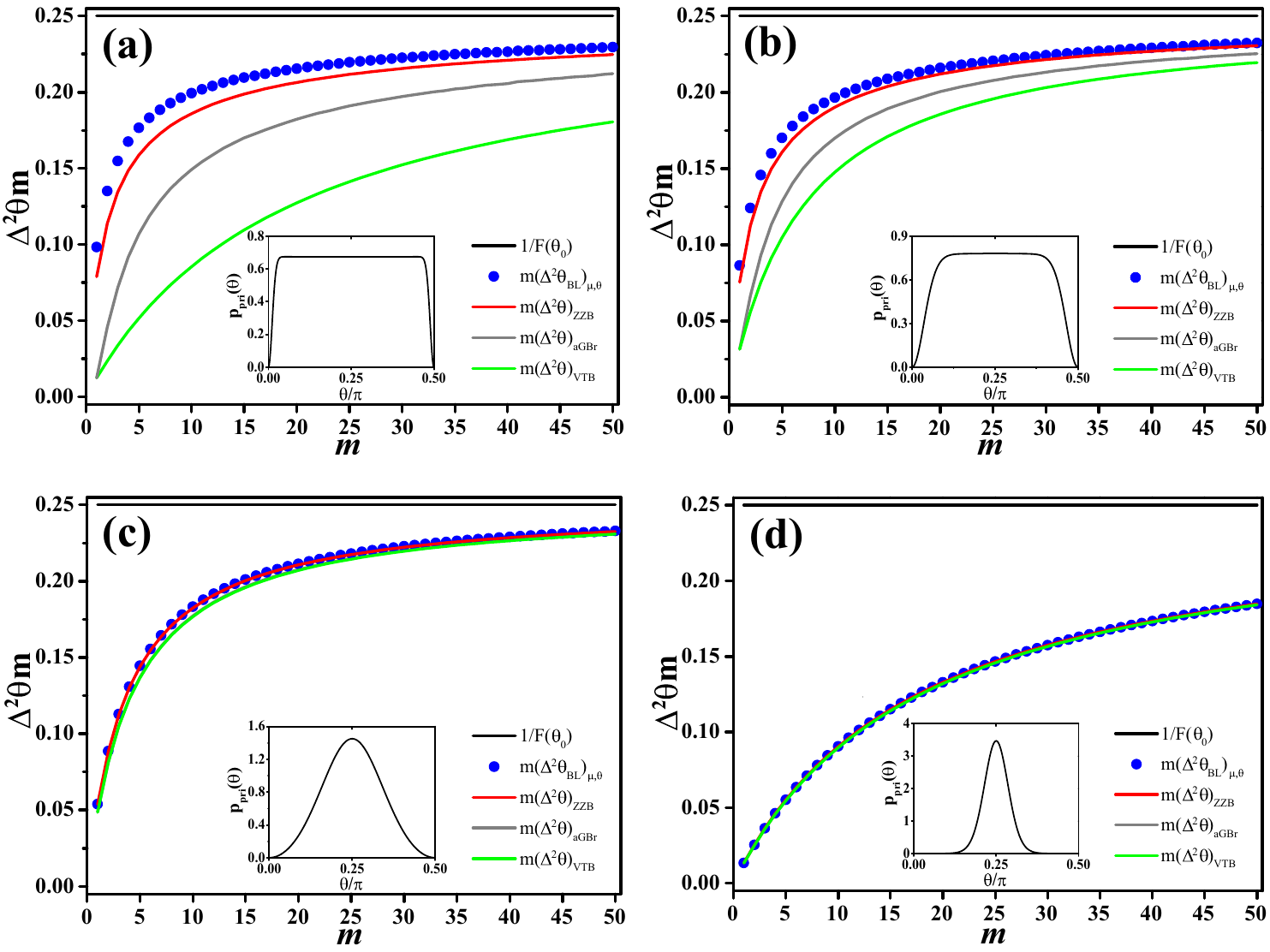}
 \caption{Average posterior Bayesian variance, $m  (\Delta^2\theta_{\rm BL} )_{\bm{\mu},\theta}$ (dots), as a function of the sample size $m$.
 Different panels corresponds to different prior distributions, as considered in Fig.~\ref{fig03}.
 This variance is compared to to the average Ghosh bound for random parameters $m(\Delta^2 \theta_{\rm aGBr})$ (grey line),
 the Van Trees bound $m(\Delta^2 \theta_{\rm VTB})$ (green line), the Ziv-Zakai bound  $m(\Delta^2 \theta_{\rm ZZB})$ (red line) and $1/F(\theta_0)$ (black horizontal line).
 }
\label{fig04}
\end{figure}

We emphasize that the average over the marginal distribution $p_{\rm mar}(\bm{\mu})$, which connects Eq.~(\ref{Bayvar}) and Eq.~(\ref{Bayvarave}),
has operational meaning if we consider that $\theta_0$ is a random variable distributed according to $p(\theta_0)$, and  $p(\theta)$ is used as prior in
the Bayes-Laplace theorem to define a posterior distribution. In this case, and under the condition $f(\bm{\mu}, a, b)=0$ (for instance if the prior distribution vanishes
at the borders of the phase domain), using Jensen's inequality, we find
\begin{eqnarray}
\Delta^2 \theta_{\rm aGBr} &=& \sum_{\bm{\mu}}  \frac{p(\bm{\mu})}
{\int_{a}^{b} d\theta \, \frac{1}{p_{\mathrm{post}}(\theta|\bm{\mu})} \big(\frac{d p_{\mathrm{post}}(\theta|\bm{\mu})}{d \theta} \big)  ^{2}}  \nonumber \\
&\geq&
\frac{1}{\sum_{\bm{\mu}} p(\bm{\mu}) \, \int_{a}^{b} d\theta \, \frac{1}{p_{\mathrm{post}}(\theta|\bm{\mu})} \big(\frac{d p_{\mathrm{post}}(\theta|\bm{\mu})}{d \theta} \big)^{2}} \nonumber \\
&=& \frac{1}{\sum_{\bm{\mu}} \int_{a}^{b} d\theta \, \frac{1}{p(\theta, \bm{\mu})} \big(\frac{\partial p(\theta, \bm{\mu})}{\partial\theta} \big)  ^{2}},
\label{avCRLBJ}
\end{eqnarray}
that coincides with the Van Trees bound discussed above. We thus find that
the averaged Ghosh bound for random parameters~(\ref{avpcrlb}) is sharper than the
Van Trees bound~(\ref{VanTrees}):
\begin{equation} \label{ineq3}
(\Delta^2 \theta_{\rm BL})_{\bm{\mu},\theta} \geq \Delta^2 \theta_{\rm aGBr} \geq \Delta^2 \theta_{\rm VTB},
\end{equation}
which is also confirmed by the numerical data shown in Fig.~\ref{fig04}.

In Fig.~\ref{fig04} we compare $\left(\Delta^{2}\theta_{\mathrm{BL}} \right)_{\bm{\mu}, \theta}$ with the various bounds
discussed in this Section. As $p(\theta_0)$ we consider the same prior~(\ref{exampleprior}) used in Fig.~\ref{fig03}. We observe that all bounds approach the Van Trees bound with increasing sharpness of the prior distribution. Asymptotically in the number of measurements $m$, all bounds converge to the Cram\`er-Rao bound.


\section{Discussion}

\begin{table}[bt]

\begin{center}
  \begin{tabular}{| c | c | c  c | c |}
    \hline
   & Risk function &  \multicolumn{2}{c|}{Bounds} & Remarks \\
    \hline
    \hline

    \multicolumn{5}{|l|}{$\theta_0$ fixed}\\[0mm]    \hline

\parbox[t]{6mm}{\multirow{4}{*}{\rotatebox[origin=c]{90}{Frequentist}}} & \multirow{2}{*}{$(\Delta^2 \theta_{\rm est})_{\bm{\mu} \vert \theta_0}$}
& BB & Eq.~(\ref{BB})&\multirow{4}{*}{hierarchy of bounds, Eq.~(\ref{ineq})} \\
 &  & EChRB & Eq.~(\ref{EXCHRB})&\\
 &  \multirow{2}{*}{$\mathrm{MSE}(\theta_{\rm est})_{\bm{\mu} \vert \theta_0}$} & ChRB& Eq.~(\ref{CHRB}) & \\
 &  & CRLB & Eq.~(\ref{CRLB}) & \\
\hline

\parbox[t]{6mm}{\multirow{4}{*}{\rotatebox[origin=c]{90}{Bayesian}}} &  \multirow{2}{*}{$(\Delta^2 \theta_{\rm BL})_{\bm{\mu} \vert \theta_0}$} & \multirow{2}{*}{GB}
& \multirow{2}{*}{Eq.~(\ref{pcrlb0})} & \multirow{2}{*}{function of $\bm{\mu}$}\\&&&&\\\cline{2-5}
 &  \multirow{2}{*}{$(\Delta^2 \theta_{\rm BL})_{\bm{\mu} ,\theta\vert \theta_0}$} & \multirow{2}{*}{aGB} & \multirow{2}{*}{Eq.~(\ref{pcrlb})}
 & \multirow{2}{*}{average over likelihood $p(\bm{\mu}|\theta_0)$}\\&&&&\\
\hline
\hline

\multicolumn{5}{|l|}{$\theta_0$ random}\\[0mm]    \hline

\parbox[t]{6mm}{\multirow{4}{*}{\rotatebox[origin=c]{90}{Frequentist}}} & \multirow{2}{*}{$(\Delta^2 \theta_{\rm est})_{\bm{\mu} , \theta_0}$} & aCRLB & Eq.~(\ref{aCRLB})
&\multirow{2}{*}{hierarchy of bounds, Eq.~(\ref{ineq2})} \\
&  & fVTB & Eq.~(\ref{VanTrees})&  \\\cline{2-5}
 & \multirow{2}{*}{$\mathrm{MSE}(\theta_{\rm est})_{\bm{\mu} , \theta_0}$} & VTB & Eq.~(\ref{VanTreesB})&\multirow{2}{*}{bounds are independent of the bias} \\
&  & ZZB & Eq.~(\ref{zivzakai})&  \\\hline
\parbox[t]{6mm}{\multirow{4}{*}{\rotatebox[origin=c]{90}{Bayesian}}} & \multirow{2}{*}{$(\Delta^2 \theta_{\rm BL})_{\bm{\mu} ,\theta, \theta_0}$} & \multirow{2}{*}{aGBr}
& \multirow{2}{*}{Eq.~(\ref{avpcrlb})}&\multirow{2}{*}{prior $p_{\mathrm{pri}}(\theta)$ and fluctuations $p(\theta_0)$ arbitrary} \\
&  &  & &  \\\cline{2-5}
 & \multirow{2}{*}{$(\Delta^2 \theta_{\rm BL})_{\bm{\mu}, \theta}$} & VTB & Eq.~(\ref{VanTreesB})&prior $p_{\mathrm{pri}}(\theta)$ and fluctuations $p(\theta_0)$ coincide \\
&  & ZZB & Eq.~(\ref{zivzakai}) & hierarchy of bounds, Eq.~(\ref{ineq3})
\\ \hline
  \end{tabular}
\end{center}

  \caption{Frequentist vs Bayesian bounds for fixed and random parameters.}
 \label{table1}
\end{table}

In this manuscript we have clarified the differences between frequentist and Bayesian approaches to phase estimation.
The two paradigms provide statistical results that have a different conceptual meaning and cannot be compared.
We have also reviewed and discussed phase sensitivity bounds in the frequentist and Bayesian frameworks, when the true value of the phase shift $\theta_0$
is fixed or fluctuates. These bounds are summarized in Table~\ref{table1}.

In the frequentist approach, for a fixed $\theta_0$, the phase sensitivity is determined from the width of the probability distribution of the estimator.
The physical content of the distribution is that, when repeating the estimation protocol, the obtained $\theta_{\mathrm{est}}(\bm{\mu})$ will fall,
with a certain confidence, in an interval
around the mean value $\langle \theta_{\rm est} \rangle_{\bm{\mu} \vert \theta_0}$ ({\it e.g.} $68 \%$ of the times 
within a $2(\Delta \theta_{\rm est})_{\bm{\mu} \vert \theta_0}$ interval for a Gaussian distribution) that, 
for unbiased estimators coincides with the true value of the phase shift.

In the Bayesian case, the posterior $p_{\mathrm{post}}(\theta\vert \bm{\mu})$ provides a degree of plausibility 
that the phase shift $\theta$ equals the interferometer phase $\theta_0$ when the data $\bm{\mu}$ was obtained.
This allows the Bayesian approach to provide statistical information for any number of measurements, even a single one.
To be sure, this is not a sign of failure or superiority of one approach with respect to the other one, since the two
frameworks manipulate conceptually different quantities. 
The experimentalist can choose to use one or both approaches, keeping
in mind the necessity to clearly state the nature of the statistical significance of the reported results.

The two predictions converge asymptotically in the limit of a large number of measurements. 
This does not mean that in this limit the significance of the two approaches is
interchangeable (it cannot be stated that in the limit of large repetition of the measurements, frequentist ad Bayesian provide the same results).
In this respect it is quite instructive to notice that the Bayesian $2 \sigma$ confidence may be below that of the Cram\'{e}r-Rao bound, 
as shown in Fig.~\ref{fig03}. This, at first sight, seems paradoxical,
since the CRLB is a theorem about the minimum error achievable
in parameter estimation theory. 
Yet, the CRLB is a frequentist bound and, again, the paradox is solved keeping in account that the frequentist and the Bayesian approaches
provide information about different quantities.

Finally, a different class of estimation problems with different precision bounds is encountered if $\theta_0$ is itself a random variable. In this case, the frequentist bounds
for the mean-square error (Van Trees, Ziv-Zakai) become independent of the bias, while those on the estimator variance are still functions of the bias.
The Van Trees and Ziv-Zakai bounds can be applied to the Bayesian paradigm if the average of the posterior variance over the marginal distribution is the relevant risk function.
This is only meaningful if the prior $p_{\rm pri}(\theta)$ that enters the Bayes-Laplace theorem coincides with the actual distribution $p(\theta_0)$ of the phase shift $\theta_0$.

We conclude with a remark regarding the so-called Heisenberg limit, which is a saturable lower bound on the CRLB over arbitrary quantum states with a fixed number of particles.
For instance, for a collection of $N$ two-level systems, the CRLB can be further bounded by $\Delta\theta_{\mathrm{est}}\geq1/\sqrt{m F(\theta_{0})}\geq 1/\big(\sqrt{m}N\big)$ \cite{Giovannetti2011, Varenna}.
This bound is often called the ultimate precision bound since no quantum state is able to achieve a tighter scaling than $N$. From the discussions presented in this article
it becomes apparent that Bayesian approaches (as discussed in Sec.~\ref{sec:Bayes}) or precision bounds for random parameters (Sec.~\ref{sec:rand}) are expected to lead to
entirely different types of `ultimate' lower bounds. Such bounds are interesting within the respective paradigm for which they are derived, but they cannot replace or
improve the Heisenberg limit since they address fundamentally different scenarios which cannot be compared in general.

\ack

This work was supported by the National Key R $\&$ D Program of China No. 2017YFA0304500, National Natural Science Foundation of
China (Grant No. 11374197), PCSIRT (Grant No. IRT13076), the Hundred Talent Program of the Shanxi Province (2018), and the Program of
State Key Laboratory of Quantum Optics and Quantum Optics Devices (No:KF201703). M. G. acknowledges support by the Alexander von Humboldt foundation.

\appendix

\section{Derivation of the Barankin bound}\label{app:Barankin}
Let $\theta_{\mathrm{est}}$ be an arbitrary estimator for $\theta$. Its mean value
\begin{equation}\label{eq:BBunbiasedcond}
\langle\theta_{\mathrm{est}}\rangle_{\boldsymbol{\mu}|\theta}=\sum_{\boldsymbol{\mu}}\theta_{\mathrm{est}}(\boldsymbol{\mu})p(\boldsymbol{\mu}|\theta)
\end{equation}
coincides with $\theta$ if and only if the estimator is unbiased (for arbitrary values of $\theta$). In the following we make no assumption about the bias of
$\theta_{\mathrm{est}}$ and therefore do not replace $\langle\theta_{\mathrm{est}}\rangle_{\boldsymbol{\mu}|\theta}$ by $\theta$.

Introducing the likelihood ratio
\begin{equation}
L(\boldsymbol{\mu}|\theta_i,\theta_0)=\frac{p(\boldsymbol{\mu}|\theta_i)}{p(\boldsymbol{\mu}|\theta_0)}
\end{equation}
under the condition $p(\boldsymbol{\mu}|\theta_0)>0$ for all $\boldsymbol{\mu}$, we obtain with Eq.~(\ref{eq:BBunbiasedcond}) that
\begin{equation}\label{eq:b4}
\sum_{\boldsymbol{\mu}}\theta_{\mathrm{est}}(\boldsymbol{\mu})L(\boldsymbol{\mu}|\theta_i,\theta_0)p(\boldsymbol{\mu}|\theta_0)=\langle\theta_{\mathrm{est}}\rangle_{\boldsymbol{\mu}|\theta_i},
\end{equation}
for an arbitrary family of phase values $\theta_1,\dots,\theta_n$ picked from the parameter domain. Furthermore, we have
\begin{equation}\label{eq:b5}
\sum_{\boldsymbol{\mu}}L(\boldsymbol{\mu}|\theta_i,\theta_0)p(\boldsymbol{\mu}|\theta_0)=\sum_{\boldsymbol{\mu}}p(\boldsymbol{\mu}|\theta_i)=1
\end{equation}
for all $\theta_i$. Multiplying both sides of Eq.~(\ref{eq:b5}) with $\langle\theta_{\mathrm{est}}\rangle_{\boldsymbol{\mu}|\theta_0}$ and subtracting it from~(\ref{eq:b4}) yields
\begin{equation}\label{eq:b6}
\fl \sum_{\boldsymbol{\mu}}\left(\theta_{\mathrm{est}}(\boldsymbol{\mu})-\langle\theta_{\mathrm{est}}\rangle_{\boldsymbol{\mu}|\theta_0}\right)
L(\boldsymbol{\mu}|\theta_i,\theta_0)p(\boldsymbol{\mu}|\theta_0)=\langle\theta_{\mathrm{est}}\rangle_{\boldsymbol{\mu}|\theta_i}-\langle\theta_{\mathrm{est}}\rangle_{\boldsymbol{\mu}|\theta_0}.
\end{equation}
Let us now pick a family of $n$ finite coefficients $a_1,\dots,a_n$. From Eq.~(\ref{eq:b6}) we obtain
\begin{equation}
\fl \sum_{\boldsymbol{\mu}}\left(\theta_{\mathrm{est}}(\boldsymbol{\mu})-\langle\theta_{\mathrm{est}}\rangle_{\boldsymbol{\mu}|\theta_0}\right)\left(\sum_{i=1}^na_iL(\boldsymbol{\mu}|\theta_i,\theta_0)\right)
p(\boldsymbol{\mu}|\theta_0)=\sum_{i=1}^na_i\left(\langle\theta_{\mathrm{est}}\rangle_{\boldsymbol{\mu}|\theta_i}-\langle\theta_{\mathrm{est}}\rangle_{\boldsymbol{\mu}|\theta_0}\right).
\end{equation}
The Cauchy-Schwarz inequality now yields
\begin{equation}
\fl \left(\sum_{i=1}^na_i\left(\langle\theta_{\mathrm{est}}\rangle_{\boldsymbol{\mu}|\theta_i}-\langle\theta_{\mathrm{est}}\rangle_{\boldsymbol{\mu}|\theta_0}\right)\right)^2\leq
\left(\Delta^2\theta_{\mathrm{est}}\right)_{\boldsymbol{\mu}|\theta_0}
\left(\sum_{\boldsymbol{\mu}}\bigg(\sum_{i=1}^na_i L(\boldsymbol{\mu}|\theta_i,\theta_0)\bigg)^2p(\boldsymbol{\mu}|\theta_0)\right),
\end{equation}
where
\begin{equation}
\left(\Delta^2\theta_{\mathrm{est}}\right)_{\boldsymbol{\mu}|\theta_0} = \sum_{\boldsymbol{\mu}}\left(\theta_{\mathrm{est}}(\boldsymbol{\mu})-
\langle \theta_{\mathrm{est}}\rangle_{\boldsymbol{\mu}|\theta_0}\right)^2p(\boldsymbol{\mu}|\theta_0)
\end{equation}
is the variance of the estimator $\theta_{\mathrm{est}}$. We thus obtain
\begin{equation}
\left(\Delta^2\theta_{\mathrm{est}}\right)_{\boldsymbol{\mu}|\theta_0}\geq \frac{\left(\sum_{i=1}^na_i\left(\langle\theta_{\mathrm{est}}\rangle_{\boldsymbol{\mu}|\theta_i}-
\langle\theta_{\mathrm{est}}\rangle_{\boldsymbol{\mu}|\theta_0}\right)\right)^2}{\sum_{\boldsymbol{\mu}}\left(\sum_{i=1}^na_iL(\boldsymbol{\mu}|\theta_i,\theta_0)\right)^2p(\boldsymbol{\mu}|\theta_0)},
\end{equation}
for all $n$, $a_i$, and $\theta_i$. The Barankin bound then follows by taking the supremum over these variables.

\section{Derivation of the Ziv-Zakai bound}
Derivations of the Ziv-Zakai bound can be found in the literature, see for instance Refs.~\cite{Bell1997,VanTreesBook2007,ZivZakai}.
This Appendix follows these derivations closely and provides additional background which may be useful for readers less familiar with the field of hypothesis testing.

Let $X\in [0,a]$ be a random variable with probability density $p(x)$.
We can formally write $p(x) = - d P(X \geq x)/dx$, where $P(X \geq x) \equiv \int_{x}^{a}p(y)dy$ is the probability that $X$ is larger or equal than $x$.
We obtain from integration by parts
\begin{eqnarray}
\langle X^2\rangle=\int_0^{a}x^2p(x)dx&=-\left[x^2P(X\geq x)\right]^{a}_0+2\int_0^aP(X\geq x)xdx \nonumber \\
&=2\int_0^{a}P(X\geq x)xdx  \nonumber \\
&=\frac{1}{2}\int_0^{2a}P\left(X\geq \frac{h}{2}\right)hdh,
\end{eqnarray}
where we assume that $a$ is finite [if $a \to \infty$ the above relation holds when $\lim_{a\to\infty} a^2P(X\geq a)=0$].
Finally, we can formally extend the above integral up to $\infty$ since $P(X\geq a) =0$:
\begin{equation}
\langle X^2\rangle=\frac{1}{2}\int_0^{\infty}P\left(X\geq \frac{h}{2}\right)hdh.
\end{equation}
Following Ref.~\cite{Bell1997}, we now
take $\epsilon=\theta_{\mathrm{est}}(\bm{\mu})- \theta_0 $ and $X  = |\epsilon|$. We thus have
\begin{equation}\label{eq:varaltexp}
\mathrm{MSE}(\theta_{\rm est})_{\bm{\mu}, \theta_0} = \langle |\epsilon|^2\rangle =\frac{1}{2}\int_0^{\infty}P\left(|\epsilon|\geq \frac{h}{2}\right)hdh.
\end{equation}
We express the probability as
\begin{eqnarray}
\fl P\left(|\epsilon|\geq \frac{h}{2}\right)&=P\left(\epsilon > \frac{h}{2}\right)+P\left(\epsilon\leq -\frac{h}{2}\right)\nonumber\\
&=P\left(\theta_{\mathrm{est}}(\bm{\mu})- \theta_0 > \frac{h}{2}\right)+P\left(\theta_{\mathrm{est}}(\bm{\mu})- \theta_0 \leq -\frac{h}{2}\right)\nonumber\\
&=\int P\Big(\theta_{\mathrm{est}}(\bm{\mu})- \theta_0 >\frac{h}{2}\Big\vert \theta_0\Big) p(\theta_0)  d \theta_0+ \nonumber \\
&\quad+\int P\Big(\theta_{\mathrm{est}}(\bm{\mu})-\theta_0 \leq -\frac{h}{2}\Big\vert \theta_0 \Big) p(\theta_0) d\theta_0.
\end{eqnarray}
Next, we replace $\theta_0$ with $\theta_0 + h$ in the second integral:
\begin{eqnarray}\label{eq:plarger}
\fl P\left(|\epsilon|\geq \frac{h}{2}\right)
&=\int P\Big(\theta_{\mathrm{est}}(x)-\theta_0 > \frac{h}{2}\Big\vert\theta_0\Big)p(\theta_0)d\theta_0+\nonumber\\&\quad+
\int P\Big(\theta_{\mathrm{est}}(x)-\theta_0 \leq \frac{h}{2}\Big\vert \theta_0 + h \Big) p(\theta_0+h) d\theta_0\nonumber\\
&=\int(p(\varphi)+p(\varphi+h))\left[\frac{p(\varphi)}{p(\varphi)+p(\varphi+h)}P\Big(\theta_{\mathrm{est}}(x)-\varphi> \frac{h}{2}\Big\vert\theta_0=\varphi\Big)\right.+\nonumber\\
&\hspace{0.5cm}+\left.\frac{p(\varphi+h)}{p(\varphi)+p(\varphi+h)}P\Big(\theta_{\mathrm{est}}(x)-\varphi\leq \frac{h}{2}\Big\vert\theta_0=\varphi+h\Big)\right]d\varphi.
\end{eqnarray}

We now take a closer look at the expression within the angular brackets and interpret it in the framework of hypothesis testing. Suppose that we try to discriminate between the
two cases $\theta_0=\varphi$ (hypothesis 1, denoted $H_1$) and $\theta_0=\varphi+h$ (denoted $H_2$). We decide between the two hypothesis $H_1$ and $H_2$ on the basis of the
measurement result $x$ using the estimator $\theta_{\mathrm{est}}(x)$. One possible strategy consists in choosing the hypothesis whose value is closest to the obtained estimator.
Hence, if $\theta_{\mathrm{est}}(x)\leq\varphi+h/2$ we assume $H_1$ to be correct and otherwise, if $\theta_{\mathrm{est}}(x)>\varphi+h/2$ we pick $H_2$.

Let us now determine the probability to make an erroneous decision using this strategy. There are two scenarios that will lead to a mistake. First, our strategy fails
whenever $\theta_{\mathrm{est}}(x)\leq\varphi+h/2$ when $\theta_0=\varphi+h$. In this case $H_2$ is true but our strategy leads us to choose $H_1$. The probability for this to
happen, given that $\theta_0=\varphi+h$, is $P(\theta_{\mathrm{est}}(x)-\varphi\leq \frac{h}{2}\vert\theta_0=\varphi+h)$. To obtain the probability error of our strategy,
we need to multiply this with the probability with which $\theta_0$ assumes the value $\varphi+h$, which is given by $p(H_2)=\frac{p(\varphi+h)}{p(\varphi)+p(\varphi+h)}$.
Second, our strategy also fails if $\theta_{\mathrm{est}}(x)>\varphi+h/2$ for $\theta_0=\varphi$. This occurs with the conditional probability
$P(\theta_{\mathrm{est}}(x)-\varphi> \frac{h}{2}\vert\theta_0=\varphi)$, and $\theta_0=\varphi$ with probability $p(H_1)=\frac{p(\varphi)}{p(\varphi)+p(\varphi+h)}$.
The total probability to make a mistake is consequently given by
\begin{eqnarray}\label{eq:perr}
\fl P_{\mathrm{err}}(\varphi,\varphi+h)&=P\Big(\theta_{\mathrm{est}}(x)-\varphi> \frac{h}{2}\Big\vert H_1\big)p(H_1)+P\Big(\theta_{\mathrm{est}}(x)-\varphi\leq \frac{h}{2}\Big\vert H_2\Big)p(H_2)\nonumber\\
&=\frac{p(\varphi)}{p(\varphi)+p(\varphi+h)}P\Big(\theta_{\mathrm{est}}(x)-\varphi> \frac{h}{2}\Big\vert\theta_0=\varphi\Big)+\nonumber\\
&\quad+\frac{p(\varphi+h)}{p(\varphi)+p(\varphi+h)}P\Big(\theta_{\mathrm{est}}(x)-\varphi\leq \frac{h}{2}\Big\vert\theta_0=\varphi+h\Big),
\end{eqnarray}
and we can rewrite Eq.~(\ref{eq:plarger}) as
\begin{equation}\label{eq:varperr}
P\left(|\epsilon|\geq \frac{h}{2}\right)=\int_{-\infty}^{\infty}(p(\varphi)+p(\varphi+h))P_{\mathrm{err}}(\varphi,\varphi+h)d\varphi.
\end{equation}

The strategy described above depends on the estimator $\theta_{\mathrm{est}}$ and may not be optimal. In general, a binary hypothesis testing strategy can be characterized
in terms of the separation of the possible values of $x$ into the two disjoint subsets $X_1$ and $X_2$ which are used to choose hypothesis $H_1$ or $H_2$, respectively.
That is, if $x\in X_1$ we pick $H_1$ and otherwise $H_2$. Since one of the two hypothesis must be true we have
\begin{eqnarray}\label{eq:eq1}
\fl 1=p(H_1)+p(H_2)&=\int_{X_1} dxp(x|H_1)p(H_1)+\int_{X_2} dxp(x|H_1)p(H_1)+\nonumber\\&\quad+\int_{X_1} dxp(x|H_2)p(H_2)+\int_{X_2} dxp(x|H_2)p(H_2)\nonumber\\
&=\int_{X_1} dxp(x|H_1)p(H_1)+\int_{X_2} dxp(x|H_2)p(H_2)+P_{\mathrm{err}}^{X_1}(H_1,H_2),
\end{eqnarray}
where the error made by such a strategy is given by
\begin{eqnarray}
\fl P_{\mathrm{err}}^{X_1}(H_1,H_2)&=P(x\in X_2|H_1)p(H_1)+P(x\in X_1|H_2)p(H_2)\nonumber\\
&=\int_{X_2} p(x|H_1)p(H_1)dx+\int_{X_1} p(x|H_2)p(H_2)dx\nonumber\\
&=p(H_1)+\int_{X_1}\left[p(x|H_2)p(H_2)- p(x|H_1)p(H_1)\right]dx.
\end{eqnarray}
This probability is minimized if $p(x|H_2)p(H_2)<p(x|H_1)p(H_1)$ for $x\in X_1$ and, consequently, $p(x|H_2)p(H_2)\geq p(x|H_1)p(H_1)$ for $x\in X_2$. This actually identifies
an optimal strategy for hypothesis testing, known as the likelihood ratio test: If the likelihood ratio $p(x|H_1)/p(x|H_2)$ is larger than the threshold value $p(H_2)/p(H_1)$
we pick $H_1$ whereas if it is smaller, we pick $H_2$. With this choice, the error probability is minimal and reads
\begin{eqnarray}
\fl P_{\min}(H_1,H_2)
&=\int_{X_2} \left[p(x|H_1)p(H_1)-p(x|H_2)p(H_2)\right]dx+\nonumber\\&\quad+\int_{X_1}\left[p(x|H_2)p(H_2)- p(x|H_1)p(H_1)\right]dx+\nonumber\\
&\quad+\int_{X_1} p(x|H_1)p(H_1)dx+\int_{X_2} p(x|H_2)p(H_2)dx\nonumber\\
&=\frac{1}{2}-\frac{1}{2}\int \left|p(x|H_1)p(H_1)-p(x|H_2)p(H_2)\right|dx,
\end{eqnarray}
where we used Eq.~(\ref{eq:eq1}).

Applied to our case, we obtain
\begin{equation}
\fl P_{\min}(\varphi,\varphi+h)=\frac{1}{2}\left(1-\sum_{\bm{\mu}} \left|\frac{p(\bm{\mu}|\theta_0=
\varphi)p(\varphi)}{p(\varphi)+p(\varphi+h)}-\frac{p(\bm{\mu}|\theta_0=\varphi+h)p(\varphi+h)}{p(\varphi)+p(\varphi+h)}\right|\right).
\end{equation}
This result represents a lower bound on $P_{\mathrm{err}}^{X_1}(\varphi,\varphi+h)$ for arbitrary choices of $X_1$. This includes the case discussed in Eq.~(\ref{eq:perr}). Thus using
\begin{equation}
P_{\mathrm{err}}(\varphi,\varphi+h)\geq P_{\min}(\varphi,\varphi+h)
\end{equation}
in Eq.~(\ref{eq:varperr}) and inserting back into Eq.~(\ref{eq:varaltexp}), we finally obtain the Ziv-Zakai bound for the mean square error:
\begin{equation}
\fl \mathrm{MSE}(\theta_{\rm est})_{\bm{\mu}, \theta_0}\geq \frac{1}{2}\int_0^{\infty} hdh\int d\theta_0(p(\theta_0)+p(\theta_0+h))P_{\mathrm{min}}(\theta_0,\theta_0+h).
\end{equation}
This bound can be further sharpened by introducing a valley-filling function \cite{BT74}, which is not considered here.

\end{document}